\documentclass[twocolumn]{aastex63}

\newcommand{\kms}{\mbox{km\,s$^{-1}$}}

\newcommand{\Msun}{\mbox{M$_{\odot}$}}

\newcommand{\E}[1]{\hbox{$10^{ #1 }$}}

\def\HII        {\hbox{H \small{II}}}

\begin{document}


\title{Magnetic Fields in Massive Star-Forming Regions (MagMaR) II. Tomography Through  Dust and Molecular Line Polarization in NGC 6334I(N)}
\shortauthors{Cort\'es et al.}

\author[0000-0002-3583-780X]{Paulo C. Cort\'es}
\affiliation{Joint ALMA Observatory, Alonso de C\'ordova 3107, Vitacura, Santiago, Chile}
\affiliation{National Radio Astronomy Observatory, 520 Edgemont Road, Charlottesville, VA 22903, USA}
\author[0000-0002-7125-7685]{Patricio Sanhueza}
\affiliation{National Astronomical Observatory of Japan, 2-21-1 Osawa, Mitaka, Tokyo 181-8588, Japan}
\affiliation{Department of Astronomical Science, SOKENDAI (The Graduate University for Advanced Studies), 2-21-1 Osawa, Mitaka, Tokyo 181-8588, Japan}
\author[0000-0003-4420-8674]{Martin Houde}
\affiliation{Department of Physics and Astronomy, The University of Western Ontario, London, ON N6A 3K7, Canada}
\author[0000-0001-9281-2919]{Sergio Mart\'in}
\affiliation{European Southern Observatory, Alonso de C\'ordova 3107, Vitacura, Santiago, Chile}
\affiliation{Joint ALMA Observatory, Alonso de C\'ordova 3107, Vitacura, Santiago, Chile}
\author[0000-0002-8975-7573]{Charles L. H. Hull}
\affiliation{National Astronomical Observatory of Japan, Alonso de C\'ordova 3788, Office 61B, 7630422, Vitacura, Santiago, Chile}
\affiliation{Joint ALMA Observatory, Alonso de C\'ordova 3107, Vitacura, Santiago, Chile}
\affiliation{NAOJ Fellow}
\author[0000-0002-3829-5591]{Josep M. Girart}
\affiliation{Institut de Ci\'encies de l'Espai (ICE-CSIC), Campus UAB, Carrer de Can Magrans S/N, E-08193 Cerdanyola del Vall\'es, Catalonia}
\affiliation{Institut d'Estudis Espacials de Catalunya (IEEC), E-08034 Barcelona, Catalonia}
\author[0000-0003-2384-6589]{Qizhou Zhang}
\affiliation{Harvard-Smithsonian Center for Astrophysics, 60 Garden Street, Cambridge, MA 02138, USA}
\author[0000-0001-5811-0454]{Manuel Fernandez-Lopez}
\affiliation{Instituto Argentino de Radioastronom\'ia (CCT-La Plata, CONICET; CICPBA), C.C. No. 5, 1894, Villa Elisa, Buenos Aires, Argentina}
\author[0000-0003-2343-7937]{Luis A. Zapata}
\affiliation{Instituto de Radioastronom\'ia  y Astrof\'isica, Universidad Nacional Autónoma de M\'exico, P.O. Box 3-72, 58090, Morelia, Michoac\'an, M\'exico}
\author[0000-0003-3017-4418]{Ian W. Stephens}
\affiliation{Department of Earth, Environment and Physics, Worcester State University, Worcester, MA 01602, USA}
\author[0000-0003-2641-9240]{Hua-bai Li}
\affiliation{Department of Physics, The Chinese University of Hong Kong, Shatin, NT, Hong Kong SAR, People's Republic of China}
\author[0000-0003-3874-7030]{Benjamin Wu}
\affiliation{NVIDIA Research, 2788 San Tomas Expressway, Santa Clara, CA 95051}
\author[0000-0002-8250-6827]{Fernando Olguin}
\affiliation{Institute of Astronomy and Department of Physics, National Tsing Hua University, Hsinchu 30013, Taiwan}
\author[0000-0003-2619-9305]{Xing Lu}
\affiliation{National
Astronomical Observatory of Japan, National Institutes of Natural Sciences, 2-21-1 Osawa, Mitaka, Tokyo 181-8588, Japan}
\author[0000-0003-0990-8990]{Andres E. Guzm\'an}
\affiliation{National Astronomical Observatory of Japan, National Institutes of Natural Sciences, 2-21-1 Osawa, Mitaka, Tokyo 181-8588, Japan}
\author[0000-0001-5431-2294]{Fumitaka Nakamura}
\affiliation{National
Astronomical Observatory of Japan, National Institutes of Natural Sciences, 2-21-1 Osawa, Mitaka, Tokyo 181-8588, Japan}
\affiliation{Department of Astronomical Science, SOKENDAI (The Graduate University for Advanced Studies), 2-21-1 Osawa, Mitaka, Tokyo 181-8588, Japan}

\email{paulo.cortes@alma.cl}

\begin{abstract}
Here, we report ALMA detections of polarized emission from dust, CS($J=5 \rightarrow 4$), and C$^{33}$S($J=5 \rightarrow 4$) toward the high-mass star-forming region NGC6334I(N). 
A clear ``hourglass'' magnetic field morphology was inferred from the  polarized dust emission which is also directly  seen from the polarized CS emission across velocity, where the polarization appears to be parallel to the field.  By considering previous findings,  the field retains a pinched shape which can be traced to clump length-scales from the envelope scales traced by ALMA, suggesting that the field is dynamically important  across multiple length-scales in  this region. 
The CS total  intensity emission is found to be optically thick ($\tau_{\mathrm{CS}} = 32 \pm 12$) while
the C$^{33}$S emission appears to be optically thin ($\tau_{\mathrm{C^{33}S}} = 0.1 \pm 0.01$). This suggests that  sources of anisotropy other than large velocity gradients, i.e. anisotropies in the radiation field are required to explain the polarized emission from CS seen by ALMA.
 By using four variants of the Davis-Chandrasekhar-Fermi technique and the angle dispersion function methods (ADF), we obtain an average of estimates for the magnetic field strength onto the plane of the sky of $\left< \mathrm{B}_{\mathrm{pos}} \right> = 16$ mG from the dust and $\left< \mathrm{B}_{\mathrm{pos}} \right> \sim 2$  mG from the CS emission, where each emission traces different molecular hydrogen number densities. This effectively enables a tomographic view of the magnetic field within a single ALMA observation. 

\end{abstract}

\section{INTRODUCTION}\label{se:intro}

Since the early days of millimeter astronomy, we have made significant progress in our understanding of 
 the physical mechanisms behind star formation in molecular clouds. 
 Because molecular clouds in the interstellar medium are composed of partially ionized gas and dust, magnetic fields are unavoidable; however, their role in formation of stars remains not well understood.
Although the past 25 years have produced significant advancement in the
 understanding of the role of magnetic fields in the star-formation process 
  \citep[see review by][]{HullZhang2019},  it is just now that new observational facilities are giving us the required
  resolution, sensitivity, and mapping capabilities that are finally allowing us to study the magnetic field in significant greater detail.

Perhaps where the role of  magnetic field is least understood is in high mass star forming  regions (HMSFR).
The main two theorized formation pathways for high mass stars, or stars with masses $\ge 8 $ {\Msun}, are either that there is a monolithic collapse from an initial massive
dense core, which is regulated by non-thermal motions or turbulence \citep{McKee2003,Krumholz2007}, or that stars form from aggregates of smaller clumps (each with an initial mass of approximately the thermal Jeans mass) that compete for gas and dust accretion and may merge to produce larger proto-stellar cores \citep{Bonnell2004,Bonnell2007}. 
Because the magnetic field has been shown to be ubiquitous in the ISM, the dynamical evolution of 
the gas and dust in high mass star-forming regions will be inevitably influenced by magnetic fields.


NGC6334 is a Giant Molecular Cloud (GMC) in the southern hemisphere. This GMC
is located inside the Sagittarius-Carina spiral arm at a distance of about $1.3 \pm 0.3$ kpc \citep{Chibueze2014} and has an estimated line mass of $\sim 1000$ \Msun\ pc$^{-1}$ with an extension of $\sim 10$ pc \citep{Andre2016}. The brightest regions
studied in the millimeter and sub-millimeter are NGC6334I and NGC6334I(N) \citep{McCutcheon2000,Hunter2014,Hunter2017,Sadaghiani2020}, where both regions appear to harbor high mass star formation. Besides the large scale mapping of polarized dust emission done by Planck \citep{Planck_Collaboration2016}, the magnetic field in NGC6334I(N) has been mapped via polarized dust emission at angular resolutions from $\sim 4^{\prime}$ \citep{Li2006}, $\sim 20^{\prime \prime}$ \cite{Vaillancourt2011}, $\sim 14^{\prime \prime}$ \citep{Arzoumanian2021}, and $\sim 2^{\prime \prime}$ \citep{Zhang2014,Li2015} which found a field shape  evolving from a  clear pinch at the high density peaks at large scales to an ``hourglass'' shape at shorter scales.   We follow the nomenclature used by \citet{HullZhang2019}, where we refer to cloud scales to structures $\sim$ 10 pc, clump scales $\sim 1$ pc, core scales between 0.1 to 0.01 pc, and envelope scales to structures $\sim$ 1000 au. In this paper, we present ALMA results of spectro-polarimetry and dust continuum polarimetry towards NGC6334I(N).
This target was observed as part of the Magnetic fields in Massive star-forming Regions (MagMaR) survey that in total contains 30 sources. Details on the survey and source selection will be given in Sanhueza et al. (2021, in prep.). Early results on specific targets are  presented in \citet{Fernandez-Lopez2021}; G5.89–0.39) and \citet{Sanhueza2021}; IRAS 18089-1732. 
The paper is organized as follows, Section \ref{se:obs} presents the observation setup including calibration and data reduction analysis, Section \ref{se:res} shows the results from polarized dust and CS emission, Section \ref{se:discussion} aggregates the analysis about the origin and ambiguities of the CS polarized emission, the magnetic field morphology and the strength estimation along with comparison to other high mass star forming regions. Finally, Section \ref{se:conc} presents the summary and conclusion from this work.

\begin{figure*} 
\includegraphics[width=1.0\hsize]{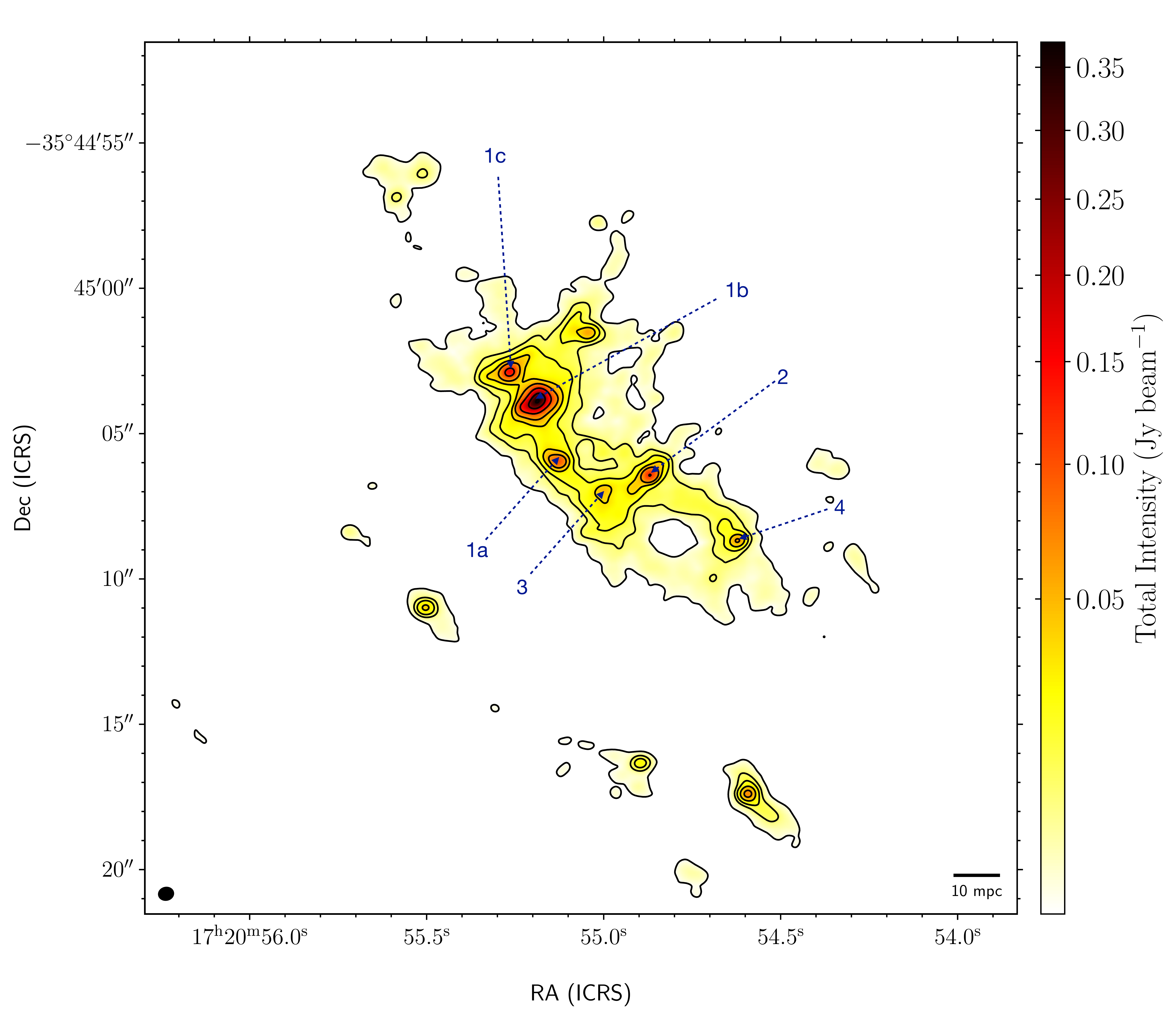}
\smallskip
\caption{ The Figure shows the total intensity dust emission at 1 mm from NGC6334I(N). The color scale is in mJy beam$^{-1}$, as indicated by the color bar. The contours correspond to the total intensity and are plotted at levels of  1.7,  8.8,  17.7, 35.3,  58.9, 117.8, 235.6, and 353.4 \,mJy\,beam$^{-1}$  with an rms for the primary beam corrected map of $\sigma = 1.1$ mJy beam$^{-1}$, where the beam is shown in the bottom-left corner as a solid, black ellipse.  The main sources as reported by \citet{Hunter2014} are indicated by the segmented arrows.
\label{fig:StkI}
}
\bigskip
\end{figure*}

\section{OBSERVATIONS}\label{se:obs}

The NGC6334I(N) source is part of project 2018.1.00105.S, which was executed twice in session mode  \citep[see chapter 8 in ][ for details about the session observing mode]{Cortes2021b}, during December 2018 and May 2019 under configuration C43-4 (providing baseline lengths from 15 to 783 m).
The correlator was configured to yield full polarization cross correlations using Frequency Division Mode  (or FDM giving $XX, XY, YX,$  and $YY$), and includes spectral windows to map the dust continuum and windows centered on major molecular line rotational transitions. 
The bandpass was calibrated using J1427-4206 for session 1 and J1924-2914 for session 2. The time dependant gain and the polarization instrumental terms were calibrated using J1717-3342 and J1751+0939, respectively.
For calibration we used CASA version 5.4 and version 5.6 for imaging \citep{McMullin2007}.
To image the continuum we manually extracted the line-free channels from each spectral window,
which we later phase-only self-calibrated using a final solution interval of 60 seconds. 
These solutions were then applied to the CS and C$^{33}$S spectral windows before imaging the lines, which were binned to 2 {\kms} per channel.  The statistics of the flat Stokes images, before debiasing, for both continuum and channel maps are shown in Table \ref{tab:stokesStats}.
All of the Stokes parameters were imaged independently using the CASA task {\em tclean}, which yielded an angular resolution of approximately $0.5^{\prime \prime} \times 0.3^{\prime \prime}$, with  a position angle of -78$^{\circ}$.
The data were primary beam corrected and debiased pixel-by-pixel following \citet{Wardle1974,Hull2015}.
Finally, we analyzed the data in the scope of the normalization issue discovered in ALMA data\footnote{See the ALMA knowledge base article at https://help.almascience.org/kb/articles/what-errors-could-originate-from-the-correlator-spectral-normalization-and-tsys-calibration}. A brief description can be found in appendix \ref{apx2}.

\section{RESULTS}\label{se:res}

\begin{figure*} 
\includegraphics[width=1.0\hsize]{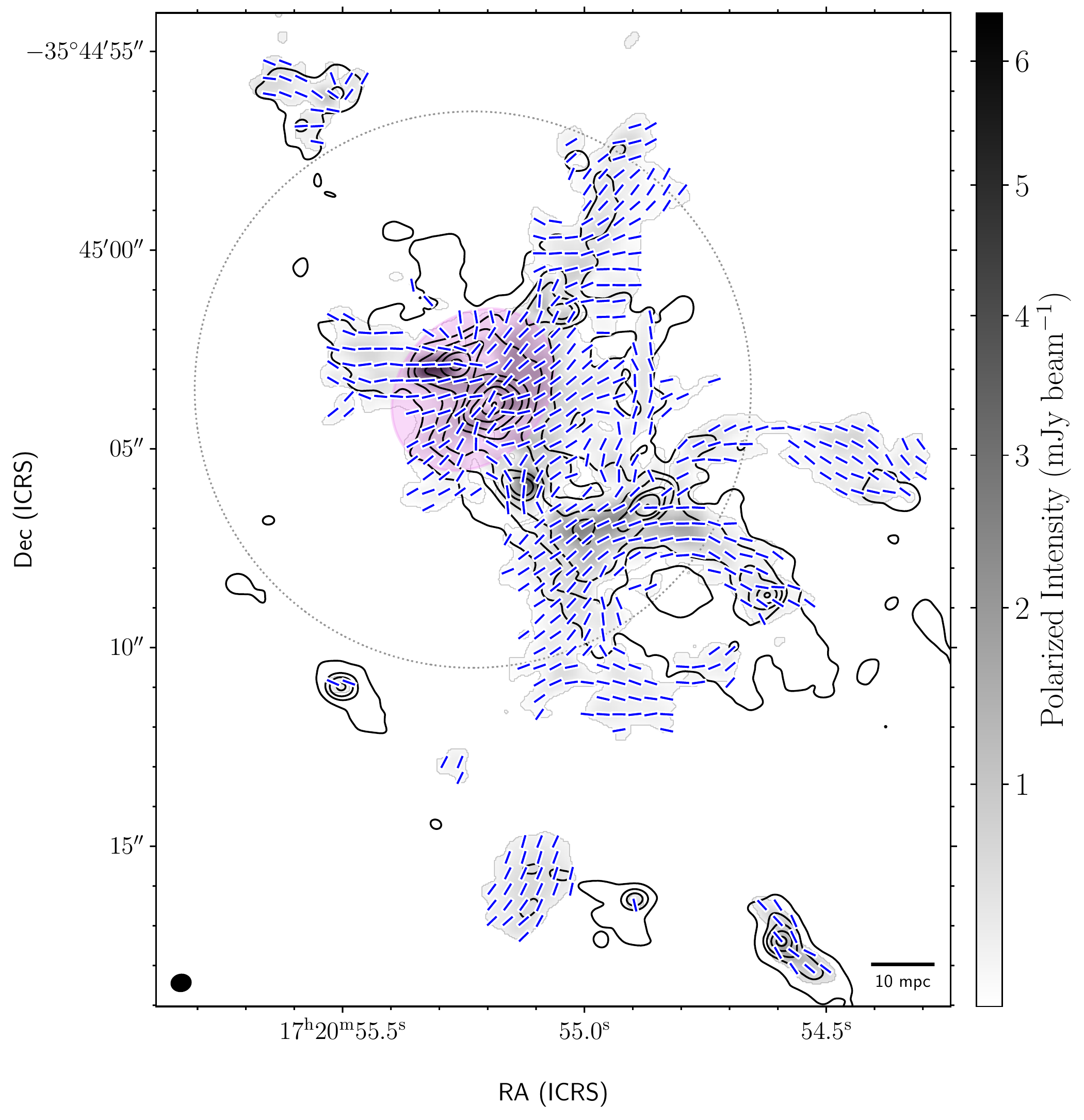}
\smallskip
\caption{The magnetic field morphology onto the plane of the sky in NGC6334I(N), as derived from the 1.3\,mm polarized dust emission is shown here. The blue line segments correspond to emission over  $3\sigma = 50\,\mu$Jy in polarized intensity after debiasing. The line segments are normalized and plotted once per synthesized beam (coarser than Nyquist sampling by a factor of two in each dimension), where the beam size is $0^{\prime \prime}.5 \times 0^{\prime \prime}.3$; the beam is shown in the bottom-left corner as a solid, black ellipse. Gray scale shows the polarized intensity in mJy beam$^{-1}$, as indicated by the color bar, and is plotted starting from 600\,$\mu$Jy beam$^{-1}$. The contours correspond to the total intensity and are plotted at levels of  1.7,  8.8,  17.7, 35.3,  58.9, 117.8, 235.6, 353.4, and 471.2 \,mJy\,beam$^{-1}$  with an rms for the primary beam corrected map of $\sigma = 1.1$ mJy beam$^{-1}$.  The purple oval encloses the 1b and 1c cores identified by \citet{Hunter2014} also  corresponding to the region used to estimate the magnetic field strength onto the plane of the sky.  The dotted circle represents the field of view of a single JCMT pointing in the data presented by \citet{Arzoumanian2021}. 
\label{fig:NGC6334IN_POL}
}
\bigskip
\end{figure*}

\subsection{Polarized Dust Continuum Emission}

\begin{figure*} 
\includegraphics[  
  width=9cm,
  height=8cm]{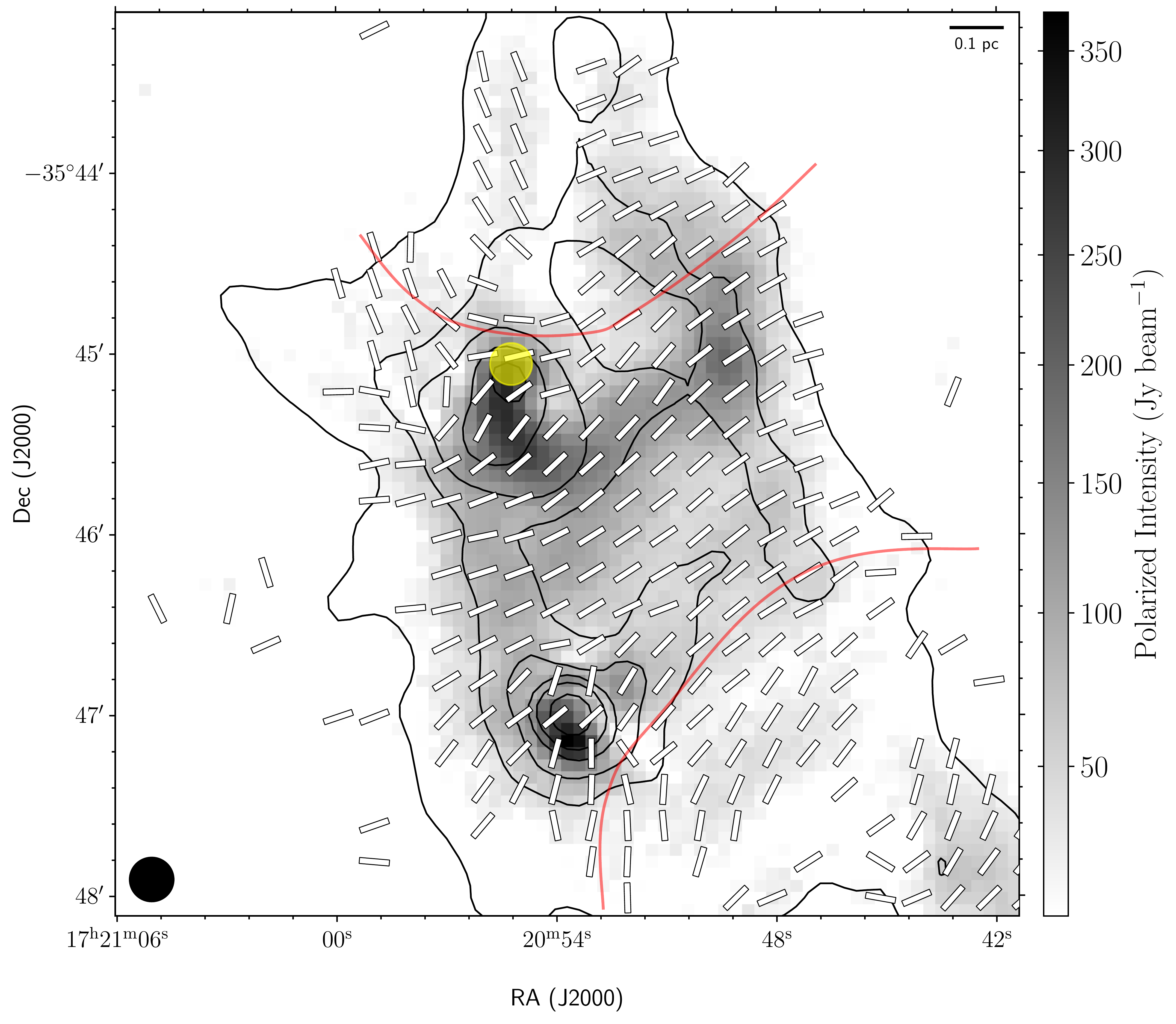}
\includegraphics[  
  width=9cm,
  height=8cm]{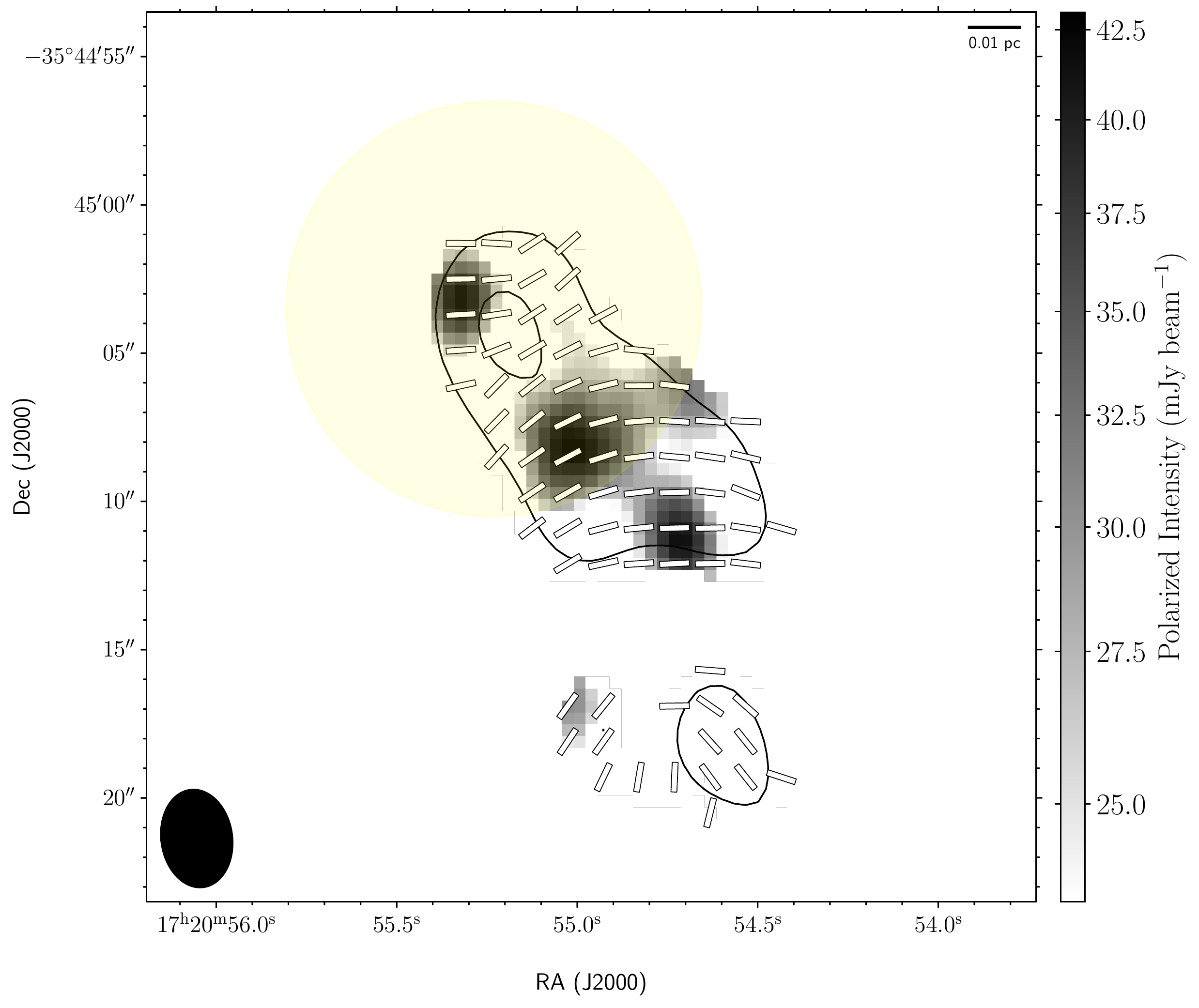}
\smallskip
\caption{The  magnetic field morphology onto the plane of the sky in NGC6334 as derived from  polarized dust continuum emission data are shown here.
{\bf\em Left.}The map shows data obtained with the JCMT at  850 $\mu$m  \citep{Arzoumanian2021}. The white line segments correspond to emission over $3\sigma = 6\,$ mJy. The line segments are normalized and plotted, approximately, once per beam, where the beam size is $14^{\prime \prime}$; the beam is shown in the bottom-left corner as a solid, black circle. Gray scale shows the polarized intensity in Jy beam$^{-1}$, as indicated by the color bar. The contours correspond to the total intensity are plotted at levels of  0.51,   2.55,   5.1,  10.2,  and 17.0 \,Jy\,beam$^{-1}$. The translucent red lines outlines the proposed ``hourglass'' pinched morphology for the mangetic field onto the plane of the sky. {\bf\em Right.} Same as the left panel, but for  SMA at 875 $\mu$m \citep{Zhang2014}. 
The white line segments correspond to emission debiased over $3\sigma$, with $\sigma \sim 8$ mJybeam$^{-1}$. The contours corresponds to step of 0.5 and 3 Jybeam$^{-1}$ and gray-scale bar indicates the polarized flux in mJybeam$^{-1}$. The SMA beam is 
2.$^{\prime\prime}1 \times 2.^{\prime\prime}07$ at a positional angle of 20$^{\circ}$.
The yellow circle shows the ALMA region of interest in both maps. 
\label{fig:jcmt}
}
\bigskip
\end{figure*}

Figure \ref{fig:StkI} shows the total intensity  (Stokes $I$) thermal dust continuum emission map from NGC6334I(N) along with the main sources identified by
\citet{Hunter2014} from their Sub Millimeter Array (SMA) data. 
The total intensity dust emission map shows an elongated filament with two cores (1b and 1c) dominating the emission and, what appears to be, a cavity in the dust emission towards the southern part of the filament. 
In this work we focus on the magnetic field leaving the cores mass and density statistical analysis for further work (Cortes et al. in prep.).
Figure \ref{fig:NGC6334IN_POL}, shows the magnetic field morphology onto the plane of the sky as derived from polarized
dust emission.
The magnetic field morphology is derived by assuming grain alignment by magnetic fields, where the polarization 
position angles are rotated by 90$^{\circ}$ to obtain the field direction.
The field pattern covers most of the NGC6334I(N) filament showing a clear indication of an ``hourglass'' shape
over the 1b and 1c  cores (inside  the purple oval in Figure \ref{fig:NGC6334IN_POL}). 
 Additionally, to the south of the main two cores we see that the field is also pinched over the third brightest core in the region (1a), with the field smoothly connecting to the aforementioned ``hourglass'' component.
A pinched field morphology in NGC6334I(N) has been suggested from cloud to core scales by \citet{Li2015}.  In fact, \citeauthor{Li2015} traced  an ``hourglass'' morphology with the SMA at core scales, which we reproduced here in Figure \ref{fig:jcmt} by using data from \citet{Zhang2014}. 
In this work we are further tracing the magnetic field morphology with ALMA from core to envelope scales. Note, we are referring to a pinched morphology instead of ``hourglass'' for the whole set of scales; we will discuss this in section \ref{sse:pattern}. 

Recently, observations with the James Clerk Maxwell Telescope (JCMT) of polarized dust emission at 850 $\mu$m revealed a detailed field morphology of NGC6334I(N) at clump scales  \citep[14$^{\prime \prime}$ resolution, ][]{Arzoumanian2021}. 
To compare with ours, we show a zoomed map of the JCMT data (see Figure \ref{fig:jcmt}, left panel), where the pinched morphology is seen over a broader region encompassing both NGC6334I(N) and NGC6334I sources (NGC6334I is not covered by the ALMA data presented here)
and outlined by translucent red lines. Over the NGC6334I(N) filament, the JCMT data shows a mostly uniform pattern covering the region mapped by ALMA (see yellow circle in Figure \ref{fig:jcmt}). Furthermore, 
\citeauthor{Arzoumanian2021} compared the JCMT data to Planck data, where pinching of the field is only seen at the North-East edge  of the cloud at the scales traced by Planck. 
Although these new data show a slightly different scenario as the one proposed by \citet{Li2015}, the magnetic field appears to evolve coherently from clump to core-envelope scales (see section \ref{sse:pattern} for a discussion). 

To the SW of the filament, a cavity is seen in the dust  emission traced by ALMA (see Figure  \ref{fig:NGC6334IN_POL}). This cavity is well 
encircled by the field, which covers most of its perimeter. Magnetic fields along cavity walls 
produced by outflows have been seen in a number of low-mass star-forming regions \citep{Hull2017b,Maury2018, LeGouellec2019, Hull2020a}. 
Coincidentally, the blue lobe of the CS outflow, previously discovered by \citet{McCutcheon2000} and also reported here, appears to be 
co-spatial with the cavity. 

To ascertain the importance of the magnetic field in NGC6334I(N), we estimate its strength onto the plane of the sky component, B$_{\mathrm{pos}}$, using a number of variants of the 
Davis, Chandrasekhar, and Fermi method \citep[or DCF:][]{Davis1951,Chandrasekhar1953,Heitsch2001,Falceta2008}, by using the
 angle dispersion function method \citep[or ADF: ][]{Hildebrand2009,Houde2009,Houde2013b,Houde2016}, and by using a recently derived approach for DCF which considers magnetosonic perturbations instead of Alfven waves \citep{Skalidis2021}. 
We will discuss the applicability of such methods to regions such as NGC6334I(N) in section \ref{sse:disp}. 
The computations were executed by following \citet{Cortes2019} for the DCF\footnote{The modifications to DCF proposed by \citet{Skalidis2021} require to change $\delta \phi$ by $\sqrt{\delta \phi}$ and a change of $1/\sqrt{2}$ scaling factor and thus the practicalities of the computation are the same as with the regular DCF variants} variants and following \citet{Houde2009,Houde2016} for the dispersion function analysis.
We obtained field-strength estimates by considering only the emission within the purple ellipse shown in Figure \ref{fig:NGC6334IN_POL} (see Table
\ref{tab:B} for the results). This is justified because is within this region  that we have obtained sufficient overlap between the polarized dust and CS emission tracing the hourglass shape of the magnetic field (see section \ref{se:cs54}).
The  B$_{\mathrm{pos}}$ estimates range between 1.4 and 23.6 mG, with an average of $\left< \mathrm{B_{pos}} \right> = 16$ mG.
In contrast with previous works, here we estimate the field strength in a self-consistent manner by using parameter values derived directly from our data.
For instance, we derive the velocity dispersion from our C$^{33}$S spectrum, which by  being optically thin (see Section \ref{se:cs54}), it traces the turbulent motions inside the  region.  The column  and  volume densities are  also derived  directly from the Stokes $I$ dust and CS emissions; we compute all values within the same region used to derive the polarization position angle dispersion ($\delta \phi$).  To derive column density from dust emission, we followed the standard approach 
\citep{Hildebrand1983} assuming a dust opacity of k$_{1.3 \textrm{mm}}$ = 0.01 cm$^{2}$ g$^{-1}$ \citep{Ossenkopf1994}, which assumes a gas to dust mass ratio of 100:1, and an average dust temperature of T$_{dust}$ = 50 K \citep{Sadaghiani2020}. We estimate the volume density by assuming a cylindrical ellipsoid as the geometrical shape of the purple ellipse in Figure \ref{fig:NGC6334IN_POL}, where the height of the cylinder is taken as the mean of the major and minor axes of the ellipse, which equals 3$\farcs$6, or 22 mpc at the distance of NGC6334I(N).
We also use a mean molecular weight of $\mu=2.8$ which assumes that the gas has a 70\% H$_{2}$ content and it is not stratified \citep{Kirk2013}.
As a result, the uncertainties in the magnetic field strength estimation result primarily from the assumptions behind the validity of the DCF method and the geometrical assumption used to compute the density. The dispersion angle analysis is discussed in Section \ref{sse:disp}.

\subsection{The polarized emission from CS ($J=5 \rightarrow 4$) and C$^{33}$S ($J=5 \rightarrow 4$)}
\label{se:cs54}

\begin{figure*} 
\centering
\includegraphics[width=\textwidth]{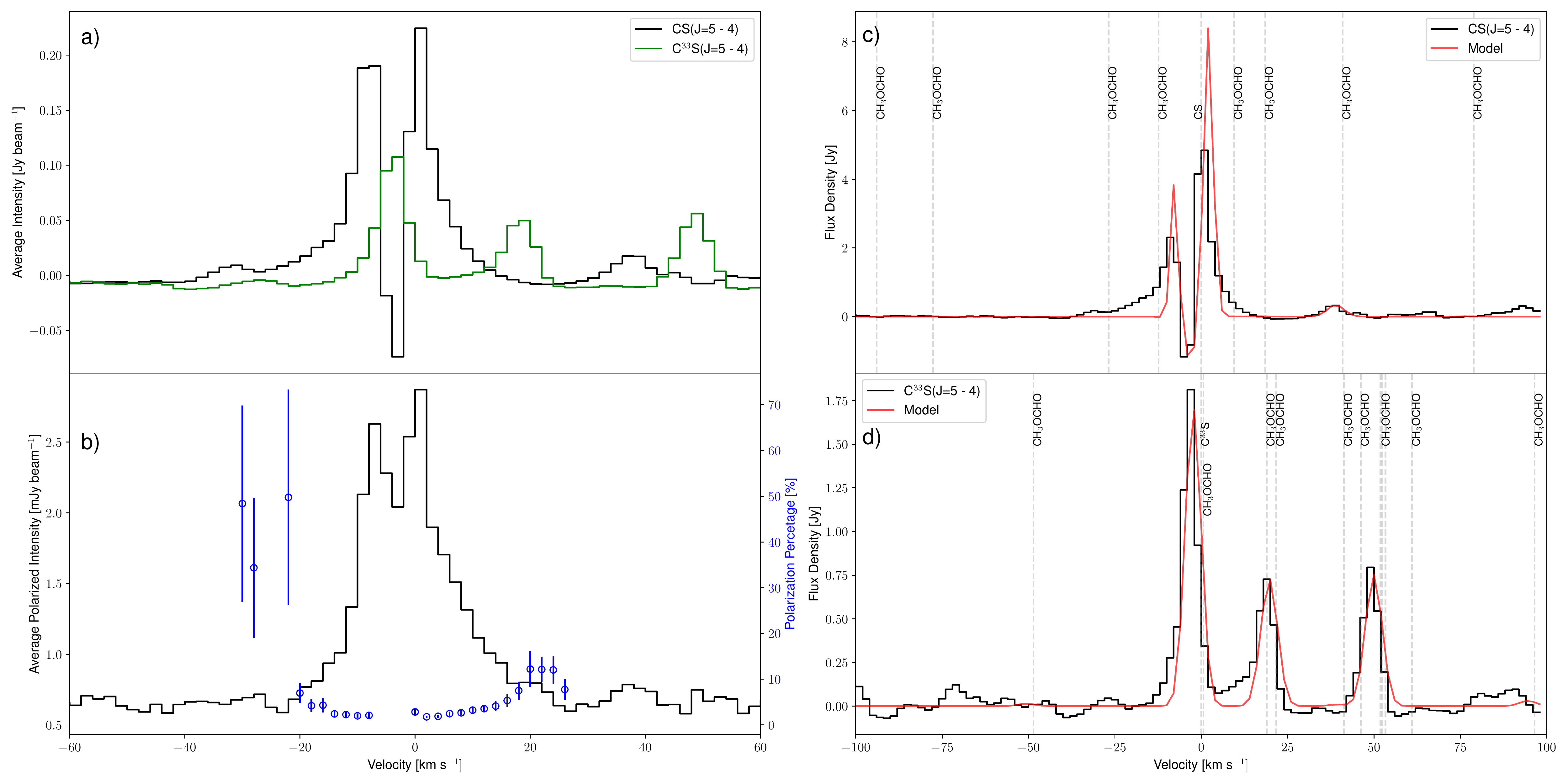}
\caption{The CS spectra from the selected  region in NGC6334I(N) is shown here (see Figure \ref{fig:NGC6334IN_POL} for the region). {\bf a)} The  panel shows the total intensity and self-absorbed CS$(J=5 \rightarrow 4)$ spectrum (blue) and the C$^{33}$S$(J=5 \rightarrow 4)$ spectrum (green). {\bf b)} The  panel shows the CS debiased polarized intensity spectrum (black) with the fractional polarization of the CS emission superposed as circles with corresponding error bars. Note, we are not showing fractional polarization values for the range between -6 to -2 {\kms} because of the negative Stokes I due to self-absorption. {\bf c)} The best fit to the CS, C$^{33}$S, and CH$_{3}$OCHO transitions is superposed on the observed  CS spectra. 
In the figure, we are also indicating additional molecular line transitions that might be present given the spectral range. We clearly detect some complex organics such as methyl formate (CH$_{3}$OCHO). {\bf d)} Same as (c) but for the C$^{33}$S emission.
The model was obtained by assuming LTE conditions, a T$_\textrm{ex} = 220$ K for C$^{33}$S, and a foreground screen of 50 K continuum to model the dust emission.
\label{fig:NGC6334IN_CS_spectra}
}
\end{figure*}

We detect CS ($J=5 \rightarrow 4$) emission, in total intensity, along the NGC6334I(N) filament across a velocity range from --40 to 20 {\kms}.  Here, we focus primarily on the polarization properties of the emission and leave a detailed analysis of the gas kinematics to a future work (Cortes et al. in prep). We analyze the CS ($J=5 \rightarrow 4$) and C$^{33}$S ($J=5 \rightarrow 4$) spectra from the same region used to estimate the magnetic field strength from polarized dust emission (see Figure \ref{fig:NGC6334IN_CS_spectra} for the spectra). 
Because C$^{33}$S is an isotoplogue of the CS molecule, we can assume that both  species are co-spatially located and thus 
we  can  use the C$^{33}$S  emission, likely  optically thin, to estimate the properties of the CS gas.
To estimate the column density and optical depth of the lines, we used the MADCUBA software package to model the CS and C$^{33}$S line profiles \citep{Martin2019};
the model is shown in Figure \ref{fig:NGC6334IN_CS_spectra}, right panel.
Under local thermodynamic equilibrium (LTE) conditions, the three
spectral features of the classic asymmetric top methyl formate (CH$_{3}$OCHO)
detected close to the C$^{33}$S line (see Figure \ref{fig:NGC6334IN_CS_spectra}) allow us to put a good
constraint to the excitation temperature, which was found to be  T$_{ex}=220 \pm 80$ K. Note, this
temperature is likely probing a gas kinetic temperature, which is significantly
higher than the assumed dust temperature, where the  CS emission probably  arises from a cooler layer  than CH$_{3}$OCHO. Although this temperature seem high, it is not uncommon to find such excitation temperatures 
when a hot molecular core (HMC) has developed, which  is  the case in   NGC6334I(N) 1b \citep{Hunter2014}. There, complex organic molecules such as methyl formate act as excellent thermometers for the gas  temperature. For instance, in W43-Main MM1 \cite{Sridharan2014} found an excitation temperature close to 400 K when considering spectral features of methyl cyanide in their data. 
Assuming that the C$^{33}$S emission is thermalized to this temperature and a source
size of $1^{\prime\prime}$ as derived from the fit to the integrated emission, we obtain a C$^{33}$S
total column density $\mathrm{N_{C^{33}S}} = 1.0  \pm 0.1 \times 10^{15}$ cm$^{-2}$, with a peak optical depth
of $\tau_{\mathrm{{C^{33}S}}} = 0.18 \pm 0.02$.
Scaling the C$^{33}$S emission using the sulphur
32/33 relative abundance ratio reported by \citet{Chin1996} yields a
CS total column density of $\mathrm{N_{CS}} = 2.9  \pm 1.1 \times 10^{17}$ cm$^{-2}$
 with a peak optical depth of $\tau_{\mathrm{CS}} = 32 \pm 12$,
 and therefore strongly optically thick. For
completeness, we modelled the CS emission with the parameters above,
together with a foreground component under T=50~K that absorbs both the
background line and continuum emission. We note the good agreement
between the velocity fit of CS, C$^{33}$S, and CH$_{3}$OCHO of $-2.46 \pm 0.15$ {\kms},
$-2.3\pm0.16$ {\kms}, and $2.0\pm0.3$ {\kms}, with the absorption layer slightly
blueshifted to $-3.4\pm0.2$ {\kms}. The  errors  are obtained from the  model fit to  the lines.

\begin{deluxetable*}{cccccccccc}
\centerwidetable
\tablecolumns{10}
\tablewidth{0pt}
\tabletypesize{\scriptsize}
\tablecaption{Statistics from Stokes Maps\label{tab:stokesStats}}
\tablehead{\colhead{Tracer} & 
           \colhead{Velocity} &
           \colhead{$I_{\mathrm{p}}\tablenotemark{\scriptsize a}$} & 
           \colhead{$\sigma_{I}$\tablenotemark{\scriptsize b}} &
           \colhead{$Q_{\mathrm{p}}$} & 
           \colhead{$\sigma_{Q}$} &
           \colhead{$U_{\mathrm{p}}$} & 
           \colhead{$\sigma_{U}$} &
           \colhead{$V_{\mathrm{p}}$} & 
           \colhead{$\sigma_{V}$} \\ 
           \colhead{} & 
           \colhead{({\kms})} &
           \colhead{(mJy beam$^{-1}$)} &
           \colhead{(mJy beam$^{-1}$)} &
           \colhead{(mJy beam$^{-1}$)} &
           \colhead{(mJy beam$^{-1}$)} &
           \colhead{(mJy beam$^{-1}$)} &
           \colhead{(mJy beam$^{-1}$)} &
           \colhead{(mJy beam$^{-1}$)} &
           \colhead{(mJy beam$^{-1}$)} 
}
\startdata
Dust & -     & 307   &  0.5  & 6.1  & 0.05 & 2.7  & 0.04 & -0.25  &  0.023 \\
CS   & -6.0  &  326  & 7   & 2.6  & 0.7 & -5.0 & 0.6 & 3.3    & 0.7 \\
CS   & -4.0  & 136   & 5   & 10.0 & 0.6 & -4.8 & 0.5 & -2.1   & 0.6  \\
CS   & -2.0  & 168   & 7   & 4.7  & 0.7 & -3.9 & 0.6 &  2.7   & 0.7
\enddata
\tablenotetext{a}{The $p$ subscript indicates peak intensity and it  applies  to  all of the Stokes parameters.}
\tablenotetext{b}{The $\sigma$ rms was estimated  by  choosing  a  region devoid of emission where the rms was obtained using the CASA task {\em  imstat}.  }
\end{deluxetable*}

\begin{figure*} 
\includegraphics[width=1.1\hsize]{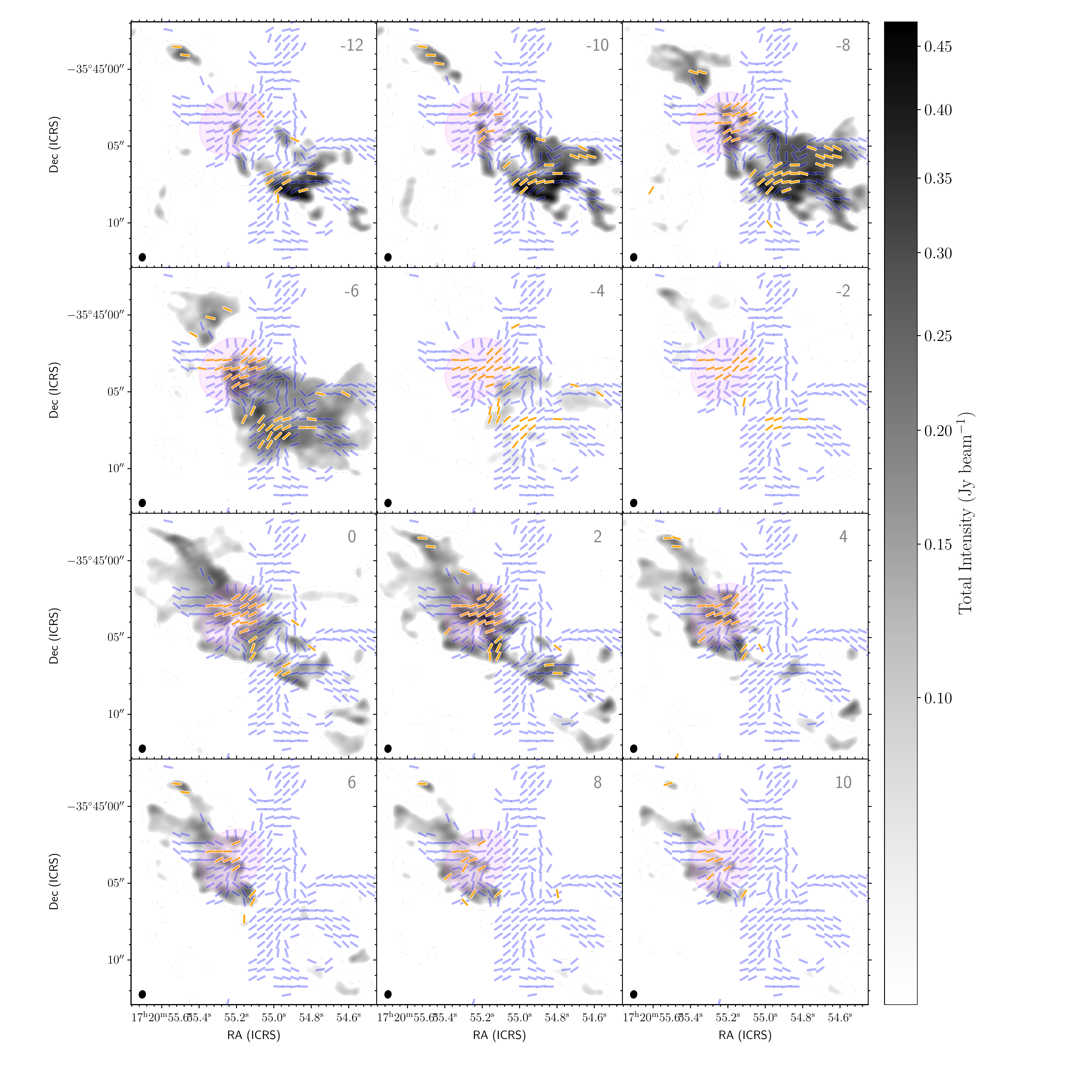}
\smallskip
\caption{Velocity channel maps of the CS$(J=5 \rightarrow 4)$ emission from NGC6334I(N) in the -12 to 10 {\kms} range. The velocity is indicated in \kms\ in the upper-right corner of each panel.  We show the Stokes $I$ CS emission in gray scale, the inferred magnetic field from polarized dust emission as blue pseudo-vectors, and the larger than $3\sigma$ significance CS polarization   angles as orange pseudo-vectors, with $\left<\sigma \right> = 780\,\mu$Jy beam$^{-1}$ is the mean rms polarized intensity emission noise when considering all channels in the range. Note,  the blue pseudo-vectors corresponding to the polarized dust emission show the inferred magnetic field morphology, having been rotated by 90$^{\circ}$ relative to the polarization. However, the polarization position angles from the polarized CS emission have not been rotated.
\label{fig:NGC6334INCS}
}
\bigskip
\end{figure*}

We detect polarized emission from the CS ($J=5 \rightarrow 4$) molecular line toward NGC6334I(N). The bottom-left panel of Figure \ref{fig:NGC6334IN_CS_spectra} shows the polarized intensity spectrum and Figure \ref{fig:NGC6334INCS} shows the channel maps. In the channel maps, we display the polarized CS emission as orange pseudo-vectors superposed on a coarsely plotted magnetic field morphology as inferred from  polarized dust emission (shown as blue pseudo-vectors). Note,  we are not applying a 90$^{\circ}$ rotation to the CS pseudo-vectors as we do to the polarized dust emission.
The channel maps reveal good agreement between the hourglass magnetic-field morphology seen in the dust and the polarized CS emission observed as a function of velocity. 
For a more quantitative comparison between the field morphology and the polarized CS emission, we compute histograms of the differences in polarization position angles between the CS and the dust emission for different velocity channels (see Figure \ref{fig:diffHist}). The differences are calculated only at locations where the polarized CS emission overlaps with the polarized dust emission. Furthermore, we fit Gaussian profiles to histograms to derive probability density functions.
All of the histograms are well centered around zero, within $\pm 10^{\circ}$ for the $\pm 4$ {\kms} range, with mostly symmetric Gaussian distributions suggesting that the CS polarized emission is correctly tracing  the ``hourglass'' field morphology derived from dust (see Table \ref{tab:csDust} for the statistics). The best match is  found at V$ = -2$ \kms\  where $\left<\Delta \theta \right> = 0^{\circ}$ and $\sigma_{\Delta \theta} = 6^{\circ}$ from the Gaussian fit.
The V$_{\mathrm{lsr}}$ of NGC6334I(N) is about  $\sim -3$ {\kms} and thus the  best match  is close to the systemic velocity of  the source.
Although there are deviations  between the magnetic field and the CS position angles, these are observed at the line-wings  of  the line suggesting a  departure from the ``hourglass'' where the emission from the outflow becomes dominant.
The polarized CS emission also seems to trace the magnetic field morphology along the dust cavity: see channel maps $v = -10$ to $-2 $ \kms,
which also show the blueshifted lobe of the CS outflow. We note that the CS outflow  appears to  be  orthogonal to the symmetry axis of the ``hourglass'' magnetic field. We  will explore this finding in  upcoming work (Cortes  et al. in prep.).
Additionally, we present channel maps of polarized emission from C$^{33}$S($J=5 \rightarrow 4$) in Figure \ref{fig:NGC6334INC33S}. Although the number of independent detections of polarization in the C$^{33}$S maps is smaller than for CS, the agreement with the CS polarization position angle appears consistent across velocity space.

The number of independent polarization detections in the CS channel maps is large enough that we can also estimate the field strength 
using the DCF technique in a number of those channels \citep[between --6 to --2 $\kms$; for an description of what can be considered to be a sufficient number of channels, see Appendix A in][]{Cortes2019}.
Over this channel range, we obtain an average field strength estimate of $\sim$ 2 mG  (see Table \ref{tab:B} for the channel-by-channel estimates). 
Polarized emission from molecular lines has a 90$^{\circ}$ ambiguity with respect to the ambient magnetic field direction \citep{Goldreich1981}. However, this does not affect the estimation method because $\delta \phi$ will not change  if the data is rotated by 90$^{\circ}$. We will address the relation between the CS polarization pattern and the magnetic field in section \ref{sse:9deg}.

\section{DISCUSSION}\label{se:discussion}

\subsection{The origin of the polarized CS emission}
\label{sse:origin}

Linearly polarized emission from molecular lines was first detected in CS emission by \citet{Glenn1997} towards the IRC +10216 evolved star.
Since then, it has been detected towards a number of sources, particularly in high-mass star forming regions \citep{Girart1999b,Lai2003a,Cortes2005,Beuther2010,Hirota2020}, and toward evolved stars \citep{Vlemmings2012,Girart2013}.
Linearly polarized emission from molecular lines is expected when the magnetic sub-levels are unevenly populated because of anisotropies in the medium. Initial models assumed that large velocity gradients resulting from the kinematics were 
the dominant source of anisotropy \citep{Goldreich1982,Deguchi1984}. However, anisotropies in the radiation field can also affect the level populations and may be due to a geometrical distribution of the gas that produces optical depths that are not the same in all directions, or from embedded sources such as a protostellar core (in the case of star-forming regions; \citealt{Cortes2005}) or a star (in the case of a circumstellar shell around evolved sources;  \citealt{Vlemmings2012}). Although it is expected that the amount of polarized emission will decrease with increasing optical depth as a result of photon trapping, we nevertheless find significant amounts of polarized emission across the CS spectrum in NGC6334I(N) (between 2\% and $\sim$ 10\% within --20 to 20 {\kms}, see the bottom-left panel of Figure \ref{fig:NGC6334IN_CS_spectra}), for which our calculations show to be optically thick.  \citet{Deguchi1984} performed multi-level calculations for the CS ($J=1\rightarrow 0$) and ($J=2\rightarrow1$) transitions and found linearly polarized emission at a level of $\sim 1$\% for $\tau \sim 10$ with a significant decrease in fractional polarization for increasing $\tau$. 
However, this computation is only one-dimensional and represents the best-case scenario where the optical depth is taken along the velocity gradient \citep[see Figure 3 in][]{Deguchi1984}. 
\citet{Cortes2005}  also performed  multilevel radiative transfer calculations for both CO ($J=1\rightarrow 0$) and ($J=2\rightarrow1$) transitions, but this time adding  a blackbody to the computation in order to introduce  anisotropies  in the radiation field.  They found increasing amount of polarization as the line becomes optically thin \citep[see   Figure 7 in ][]{Cortes2005}. 
Recently, \citet{Lankhaar2020} presented a comprehensive quantum mechanical treatment of the alignment of the molecule's angular momentum by assuming only an anisotropic radiation field. Their treatment is fully three-dimensional, allowing for simulations of both radiative transfer and gas dynamics. Although they did not model CS emission in particular, their simple example of a collapsing spherical cloud produced polarization fractions for HCO$^{+}$ ($J=3 \rightarrow 2$) and ($J=2 \rightarrow 1$) at levels over  1\% for radial distances beyond  600 au  ($\sim$ 3 mpc) for  the ($J=3 \rightarrow 2$) and 900 au ($\sim$ 4.5 mpc) for the ($J=2 \rightarrow 1$) transition. However, all of these calculations find  that polarized line  emission significantly decreases  with higher optical depths. A possible explanation to this  may  be   related to interferometric filtering and missing flux. Fractional polarization with interferometers has to be analyzed with care because of spatial filtering effects \citep{LeGouellec2020}. The maximum recoverable angular scale for the configuration used to acquire these data is  $\sim 5^{\prime\prime}$, which suggests that a significant fraction of the  extended total  intensity CS emission seen from the single dish \citep{McCutcheon2000} might be filtered-out by  ALMA and thus the fractional polarization values presented here are over-estimated. This is because linearly polarized emission tend to be  more compact than the total intensity emission. 
Estimating how far  we are from the true fractional  polarization  will require sampling larger angular scales in full polarization mode, which the ALMA compact array (ACA) can do.

We also see significant amounts of fractional polarization at higher velocities (polarization levels $\gtrsim 10\%$ at $v < -20$ and 
$v > 20$ {\kms}). It is likely that, in these velocity ranges where the CS emission is tracing the outflow, the CS emission might be optically thin and thus the values estimated  might be closer to the ``true'' fractional polarization values. The large velocity gradient and the radiation field from the embedded protostars are, most likely, the sources of anisotropy necessary to produce the large polarization fractions that we see there \citep[see][Figure 7 for a CO computation]{Cortes2005}. We also note that strong polarization from CS is expected because of its large dipole moment ($\mu =$ 1.96 D) compared with, for example, that of CO ($\mu =$0.12 D), because of the $\mu^{2}$ dependence of the radiative rates. 
Thus, it seems that the combination of anisotropies in both the radiation field and the gas kinematics is causing the strong CS polarized emission that we see in NGC6334I(N), where  filtering  effect might explain  the large fractional  polarization  values  seen when comparing to radiative transfer calculations. Nonetheless, 3-dimensional modeling of polarized CS emission will be needed to further understand the emission detected here, which is beyond the scope of this work.

Finally, if a foreground screen of molecular gas is present between the source and the telescope, some amount of linear polarization might be converted into circular polarization through anisotropic resonant scattering, or ARS \citep{Houde2013a}. The ARS will systematically corrupt the linear polarization position angle from the CS emission and change its relation respect to the ambient magnetic field. The ARS will manifest itself by the presence of statistical significant signal in Stokes $V$, which we do not detect. The peak emission is Stokes $V$ is about between -2.1 and 3.3  mJy beam$^{-1}$  with a fractional level between  0.6 to 1.0\% which is below the statistical uncertainty of $\sim$ 2\% that ALMA  can  measure \citep{Cortes2021b}.
The morphology of the Stokes $V$ velocity channel maps are consistent with noise where the  peaks alternate between positive and  negative  between  channels  which is inconsistent with a coherent  conversion of linear to circular polarization by ARS (see Table \ref{tab:stokesStats} and Figure \ref{fig:StokesV} in Appendix \ref{apx1} for the Stokes $V$ channel maps).
Thus, we conclude that polarized emission from CS is purely linear and its relation to the field is subject to the known 90$^{\circ}$ ambiguity (see next section for a discussion).

\begin{deluxetable*}{c c c c c}
\centerwidetable
\tablewidth{0pt}
\tablecolumns{5}
\tablecaption{Polarization Angle Statistics
\label{tab:csDust}}
\tablehead{\colhead{Velocity} & 
           \colhead{$\left< \Delta \phi \right>$\tablenotemark{\scriptsize a}} & 
           \colhead{$\mathrm{std}(\Delta \phi)$\tablenotemark{\scriptsize b}} & 
           \colhead{$\Delta \phi_{0}$\tablenotemark{\scriptsize c}} & 
           \colhead{$\sigma_{\Delta \phi}$\tablenotemark{\scriptsize d}}  \\ 
           \colhead{(km s$^{-1}$)} & 
           \colhead{($^{\circ}$)} & 
           \colhead{($^{\circ}$)} &
           \colhead{($^{\circ}$)} &
           \colhead{($^{\circ}$)} 
}
\startdata
-12.0          &   8 & 25.9           &  21            & -11.5\\
-10.0          &   9 & 20.1           &  16            & \phd17.4\\
-8.0           &  10 & 18.0           &  11            & -17.4\\
-6.0           &   4 & 13.8           &   \phantom{0}4 & -13.1\\
-4.0           &   2 & \phantom{0}9.0 &   \phantom{0}1 & \phd-8.4\\
-2.0           &   1 & \phantom{0}7.5 &   \phantom{0}0 & \phd6.4\\
\phantom{0}0.0 &   3 & 12.3           &   2            & \phd11.5\\
\phantom{0}2.0 &   0 & 14.4           &  -2            &  \phd7.6\\
\phantom{0}4.0 &  -2 & 14.8           &  -5            & \phd-10.1\\
\phantom{0}6.0 &  -2 & 15.6           &  -5            & -7.4\\
\phantom{0}8.0 &  -8 & 19.0           &  -6            & \phd6.2\\
10.0           &  -8 & 17.4           &  -8            & \phd6.9\\
12.0           &  -2 & 13.5           &  -4            & \phd8.7\\
14.0           &  -2 & 12.2           &  -7            & \phd8.0
\enddata
\tablenotetext{a}{Here $\left< \Delta \phi \right>$ represents the mean of the difference  values between dust and CS polarization angles.}
\tablenotetext{b}{Here $\mathrm{std}(\Delta \phi)$  corresponds to the standard deviation of the difference values between the dust and the CS polarization angles.}
\tablenotetext{c}{Here $\Delta \phi_{0}$ corresponds to the center of a Gaussian fit to the distribution of polarization angle differences.}
\tablenotetext{d}{Here $\sigma_{\Delta \phi}$ corresponds to the width of a Gaussian fit to the distribution of polarization angle differences.}
\end{deluxetable*}

\begin{figure*} 
\includegraphics[width=0.5\hsize]{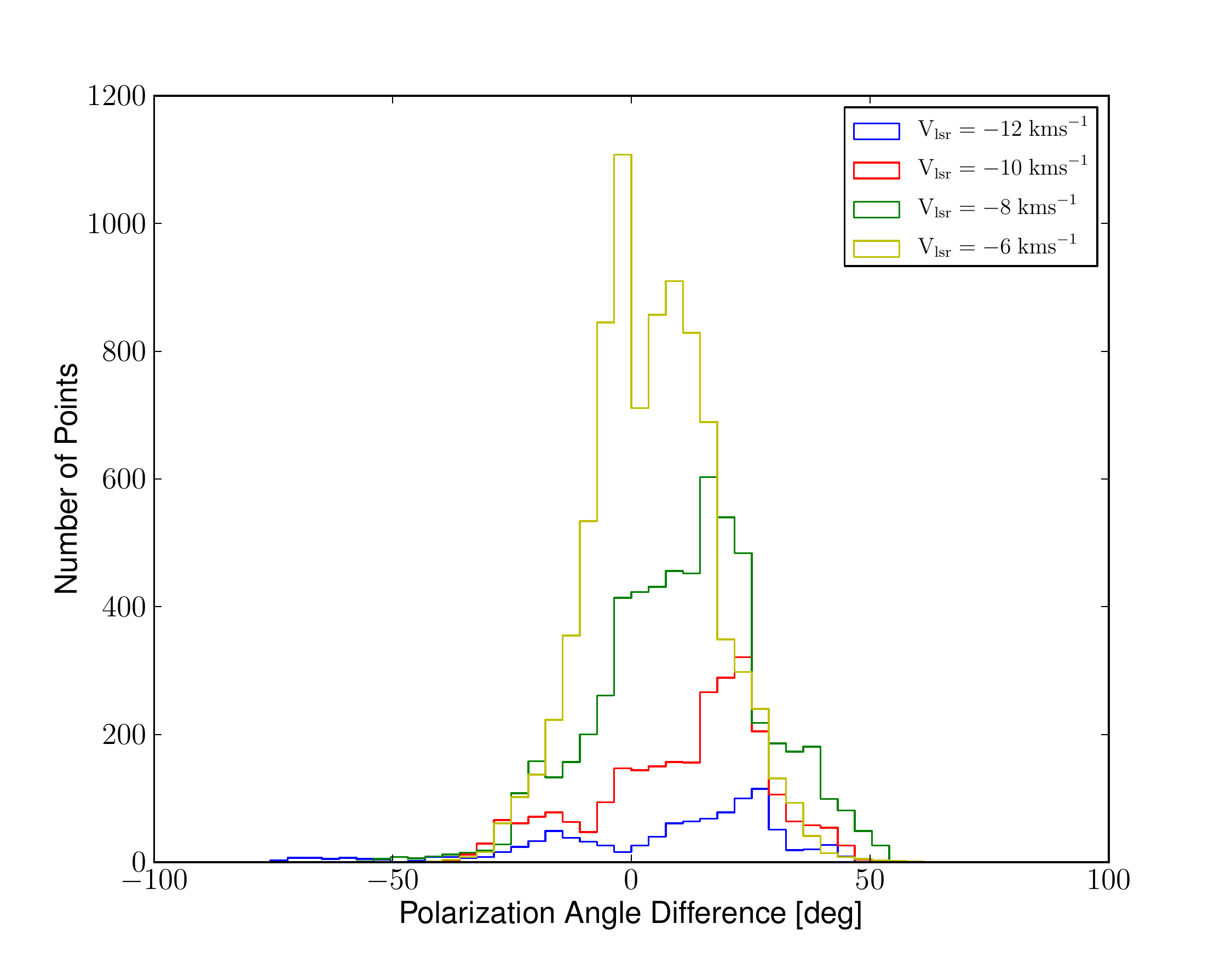}
\includegraphics[width=0.5\hsize]{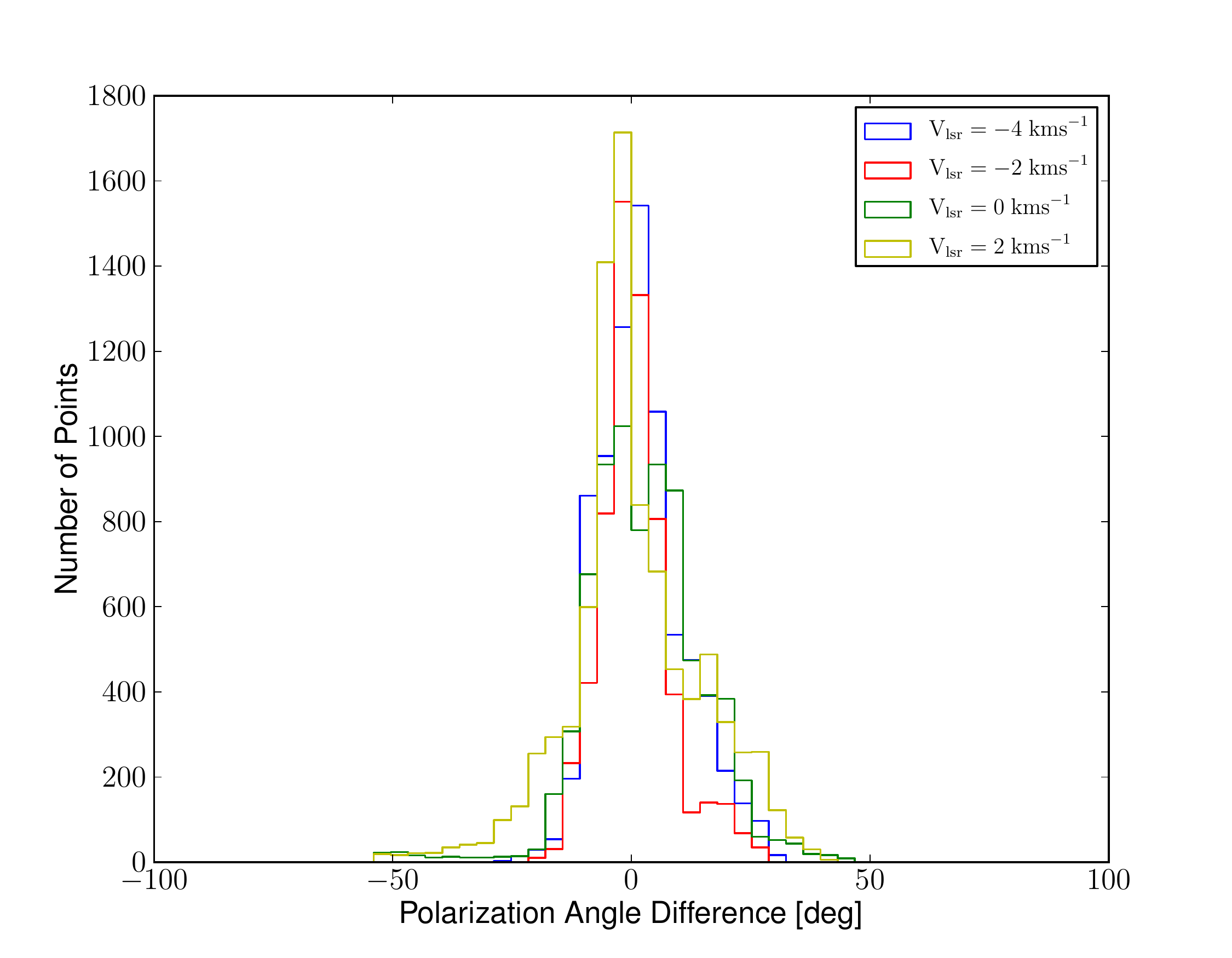}
\includegraphics[width=0.5\hsize]{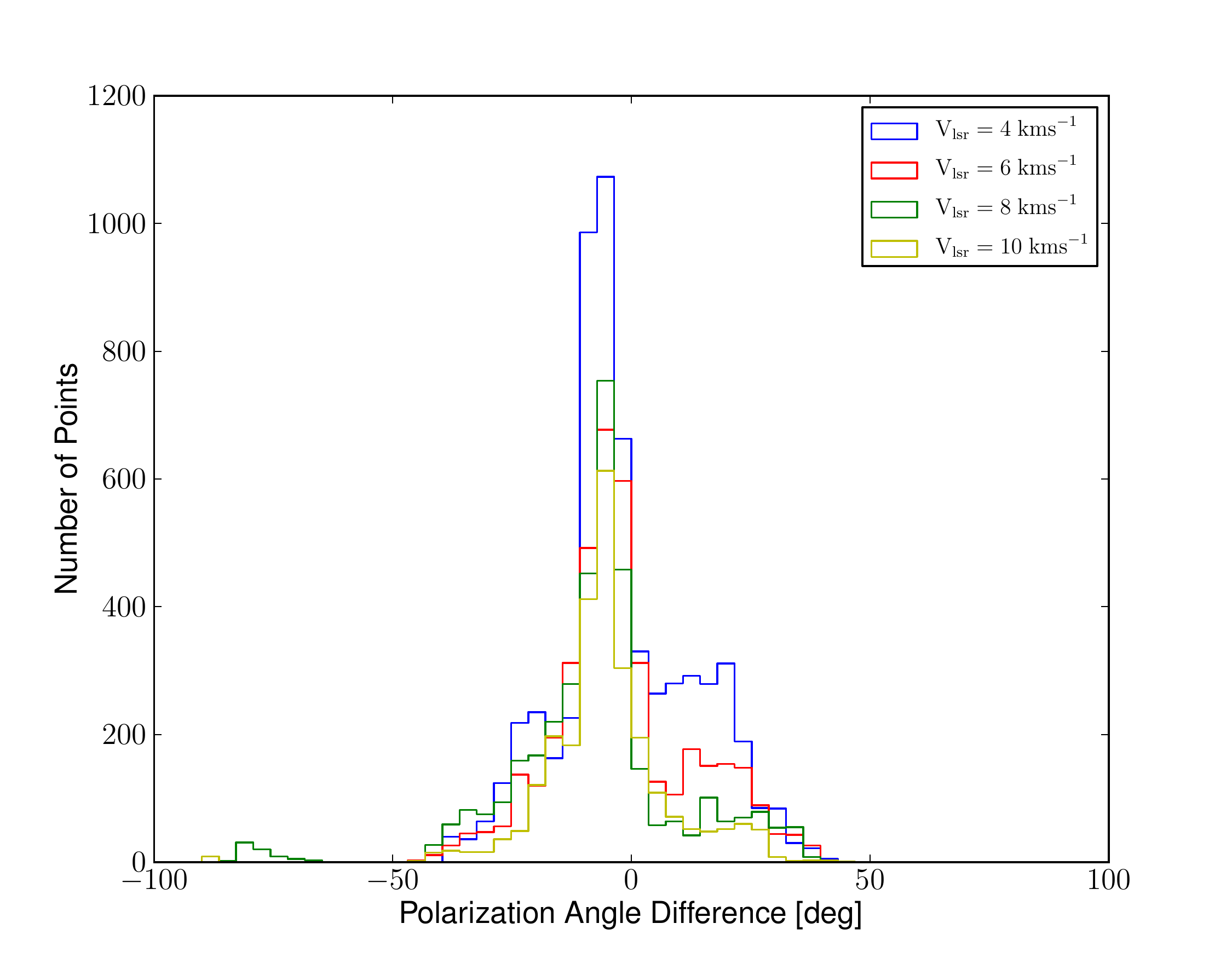}
\includegraphics[width=0.5\hsize]{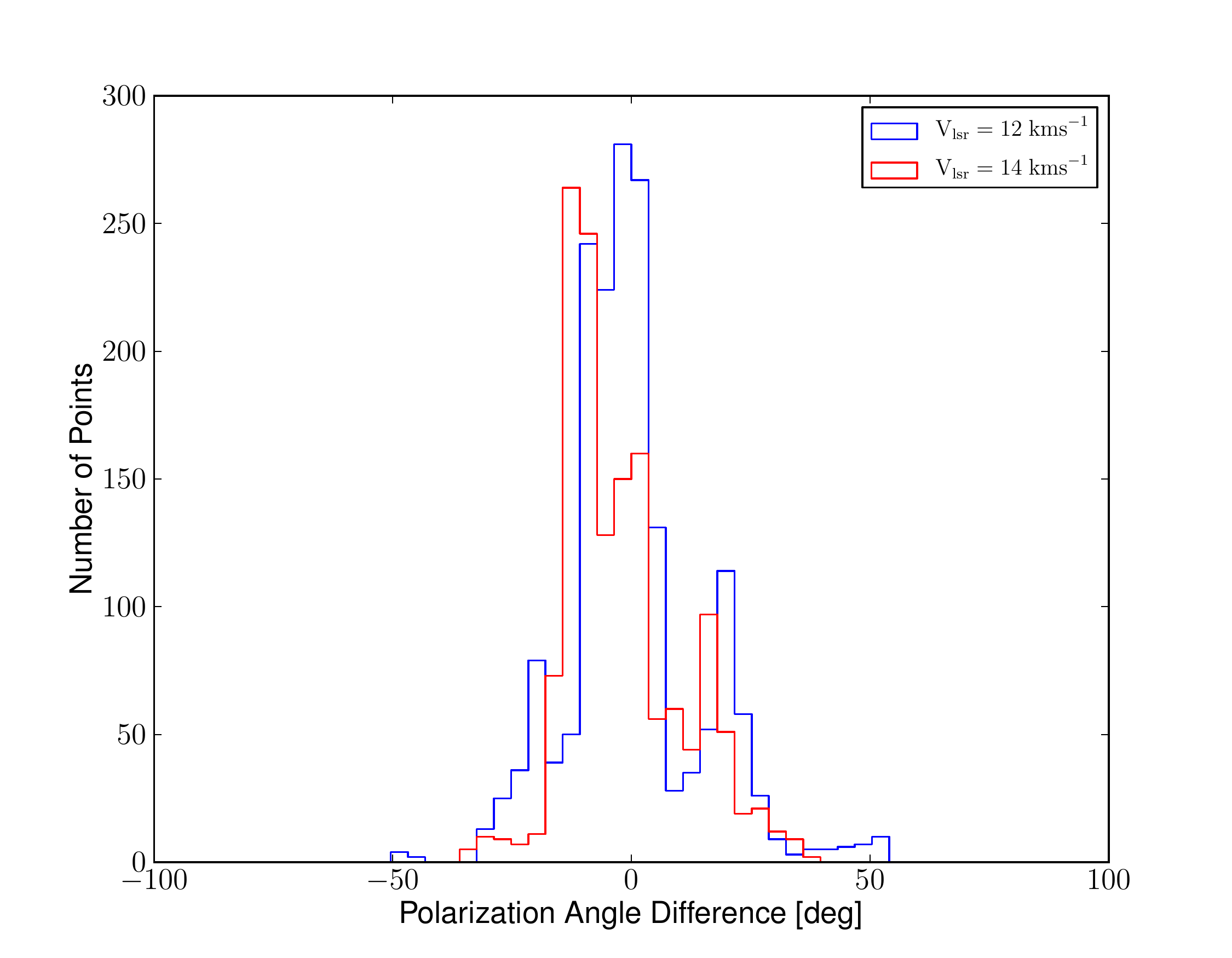}
\includegraphics[width=0.5\hsize]{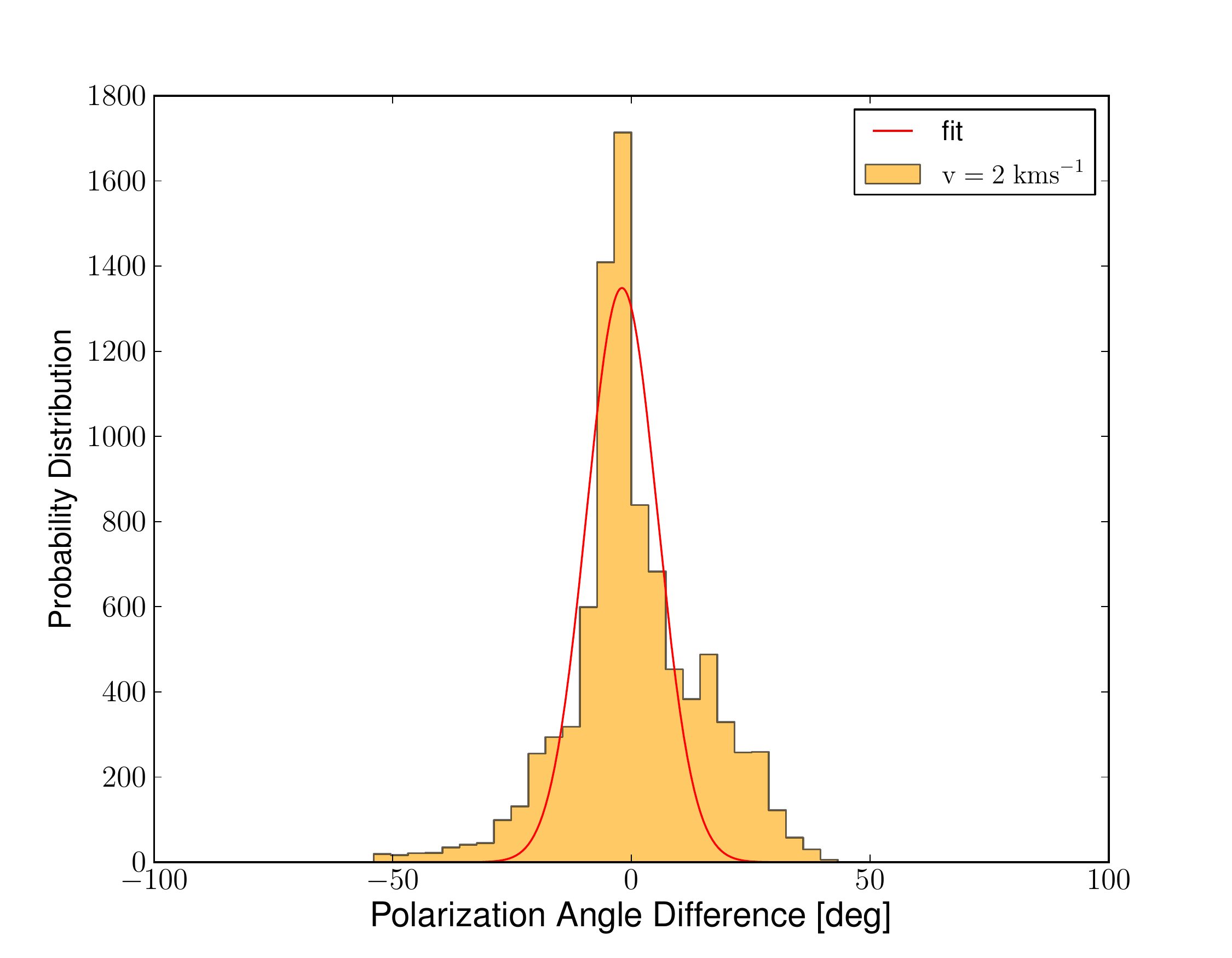}
\includegraphics[width=0.5\hsize]{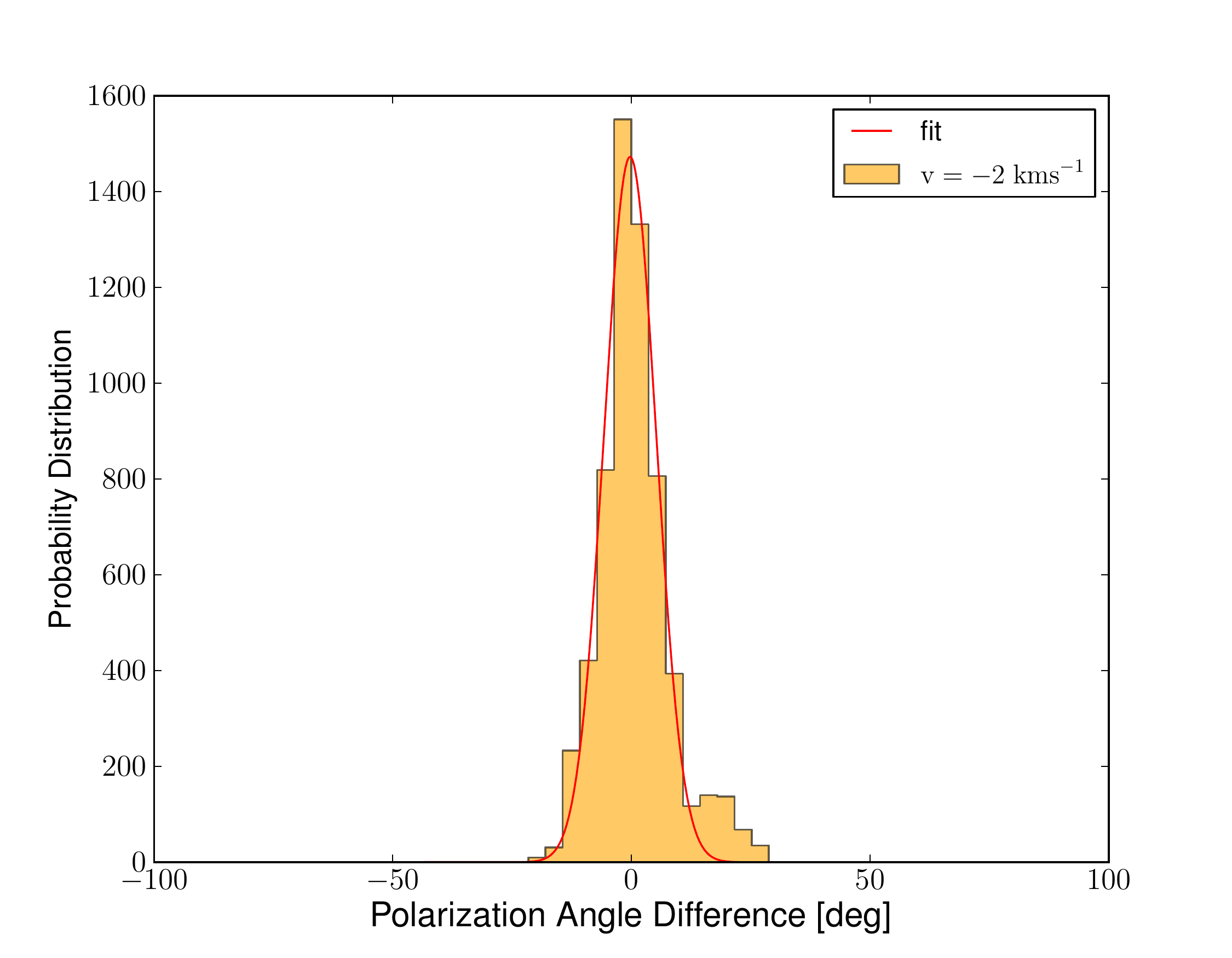}
\smallskip
\caption{Histograms of differences in the polarization position angles of the dust and CS emission.
We compute the differences inside the region used to estimate the field strength (see the purple ellipse in Figure \ref{fig:NGC6334IN_POL}). The panels are ordered as a function of velocity (from upper-left to lower-right), as indicated in legend in each plot. The best agreement is found at -2 and 2 {\kms} where Gaussian fits to the normalized histograms are shown in the two bottom panels. The fit parameters are listed in Table \ref{fig:diffHist}.
\label{fig:diffHist}
}
\bigskip
\end{figure*}

\subsubsection{The 90$^{\circ}$ ambiguity in the position angle of the polarized CS emission}
\label{sse:9deg}

Linearly polarized emission from molecular lines has a 90$^{\circ}$ ambiguity in the orientation of the polarization position angle with respect to the ambient magnetic field direction \citep{Goldreich1981}.
Thus, deriving the magnetic field morphology from the polarized CS emission has an additional degree of complexity versus the more straightforward method of deriving the field morphology from polarized dust emission.
To interpret the relationship between the CS polarization angle and the orientation of the ambient magnetic field, we use the inferred field morphology from the dust emission and arguments about physical plausibility. We interpret the position angle from the polarized CS emission as being parallel to the magnetic field onto the plane of the sky because of the close agreement between the hourglass magnetic field morphology seen in the dust and the polarization pattern of the CS emission, as shown in Figure \ref{fig:diffHist} and Table \ref{tab:csDust}. The opposite case, i.e., where the CS polarization is perpendicular to the inferred magnetic field from the dust, would imply an unrealistically complex field morphology over the central region of NGC6334I(N): because of the close agreement between CS and dust, the twisting of the field lines would have to be perpendicular at almost all positions and velocities shown in the channel maps. Such a situation is unheard of and likely nonphysical, even under strong rotation, which we do not have evidence for at the scales traced by the CS
polarized emission. 
Our interpretation that the CS polarization traces the inferred magnetic field shape is also supported by the C$^{33}$S($J=5 \rightarrow 4$) results, which agree well with the CS polarization, particularly at velocities of --4 and --2 {\kms}. These three independent polarization tracers show essentially the same result, which is a magnetic field morphology with an ``hourglass'' shape over the main cores in NGC6334I(N).


\subsection{Tomography of magnetic fields}
\label{sse:pattern}

Hourglass magnetic field morphologies have long been predicted by magnetically regulated star formation models \citep[e.g.,][]{Mouschovias1976,Mouschovias1985}, and may be the result of ambipolar diffusion \citep{Mouschovias1991b}. This field morphology has been seen in a number of low- and high-mass star forming regions and at different length and mass scales \citep{Schleuning1998,Girart2006,Girart2009,Maury2018,Beltran2019}. In the magnetically regulated star formation scenario, a molecular cloud will initially be supported against gravitational collapse by the magnetic field, which is initially thought to be uniform. Because the neutrals are only weakly coupled to the charge carriers, which are held in place by the field, they will slowly diffuse past the magnetic field via ambipolar-diffusion.  Because the field is frozen into the charge carriers, the field will be pinched attaining the hourglass shape. Alternatively, a self-gravitating core can also pull the field at the core center which will create an ``hourglass'' shape, but in this case ambipolar-diffusion might be a by-product of the process.

Although the ``hourglass'' shape seen in NGC6334I(N) is clear, the exquisite sensitivity and resolution of the ALMA data allow us 
to see local deviations from which the ``hourglass'' is a simple model. The density regime traced by ALMA is likely showing the result of complex physics which is difficult to explain without detailed numerical modeling, outside the scope of this paper. Furthermore, a clear ``hourglass'' field morphology in NGC6334I(N) is not seen at all length-scales. From the envelope scales traced by ALMA, to the core scales imaged by the SMA, and to the clump scales traced by the JCMT, the hourglass appears to be the shape of the magnetic field when we also consider  NGC6334I  (see Figures \ref{fig:NGC6334IN_POL} and \ref{fig:jcmt}) . However, at the cloud scales traced by 
Planck \citep[see Figure 5 in ][]{Arzoumanian2021}, the field is only pinched at the North-East side of the cloud. Whether this is a projection effect or not is not clear from the Figure alone where a detailed analysis of the Planck data would be required. 
Even though we cannot state with certainty that we see an ``hourglass'' field morphology as the ubiquitous shape across all length scales in NGC6334, the pinching in the field seen from cloud to the envelope scales is quite remarkable suggesting that the magnetic field is strong in this region. Furthermore, 
the field pattern seems also preserved in velocity space from --10 to 4 {\kms} at the ALMA scales, which is a 14 {\kms} range, almost three times the 5.3 {\kms} full width at half maximum (FWHM) line-width of C$^{33}$S  (our proxy for the turbulent motions). 

Besides tracing a pinched field shape through multiple orders of magnitude in spatial scales,  we are also tracing the field at different densities as well \citep[from 10 to 10$^{7}$ cm$^{-3}$ when considering density estimates from ][]{Li2015, Arzoumanian2021}. 
The critical density for the CS ($J=5 \rightarrow 4$) transition is $9 \times 10^{6}$ cm$^{-3}$, which is obtained by assuming that the emission is optically thin. Because of the high optical depth estimated for the CS ($J=5 \rightarrow 4$) transition, that number density is most likely smaller as the line may be sub-thermally excited. 
\citet{Shirley2015} accounted for optical depth effects by computing the critical density considering photon trapping. Thus, the number density can be approximated to $n_{crit}^{thick} = n_{crit}^{thin}/\tau_{\nu_{jk}}$ when the optical depth is much larger than one, where $n_{crit}^{thick}$ is the optically thin critical density and $\tau_{\nu_{jk}}$ is the line optical depth. From this assumption we obtain $n_{\mathrm{CS}, crit}^{thick} = 2.8 \times 10^{5}$ cm$^{-3}$. Note, here the $n_{\mathrm{CS}, crit}^{thick}$ refers to the collisional partners involve in the excitation of the CS molecule which corresponds to H$_{2}$. Although an abuse of notation, we use CS in the subscript to indicate that the density is derived from the CS emission.
Furthermore, \citet{Shirley2015}
tabulated effective excitation  densities for the CS molecule by considering a number of rotational transitions. The effective excitation density is an empirical quantity defined by considering a 1 K\,\kms\ integrated emission. This quantity also takes into account optical depth effects such as radiation trapping \citep[for a review, see][]{Evans1999}. 
For the CS ($J=5 \rightarrow 4$) transition, \citet{Shirley2015} obtains $n_{\mathrm{eff}} = 7.6\times10^{4}$ cm$^{-3}$, assuming a kinetic temperature of 50 K, the dust temperature which is a good estimation of the kinetic temperature in the region of interest (see purple oval in Figures \ref{fig:NGC6334IN_POL} and \ref{fig:NGC6334INCS}). Although these two estimates of the number density are close in value, detailed numerical radiative transfer modeling is required to obtained more accurate population numbers which we leave for future work. Nonetheless, the two criteria are reasonable approximations to the number density of the molecular hydrogen CS collisional partners under high optical depth conditions and thus we use the average as the estimate  of the ``true'' number density, or $n_{\mathrm{CS}} \sim 2\times10^{5}$ cm$^{-3}$.

A value of $n_{\mathrm{CS}} = 2\times10^{5}$ cm$^{-3}$ is $\sim$ 2 orders of magnitude less than the volume density estimated from the dust emission, which we calculated to be 4.2 $\times 10^{7}$ cm$^{-3}$. Thus, the polarized CS emission appears to be tracing the field at lower densities than the dust and at a level comparable with the JCMT observations. Therefore, within a single ALMA data-set we are not only tracing the ``hourglass'' shape as function of velocity, but also as a function of density: i.e., we are effectively performing magnetic field tomography. When we consider the evolution of the field morphology along these three axes (length-scale, velocity, and density), the striking coherence seen in the field structure strongly suggests that the magnetic field remains dynamically important from the diffuse to the high-density regime in NGC6334I(N).

\begin{figure*} 
\includegraphics[width=1.1\hsize]{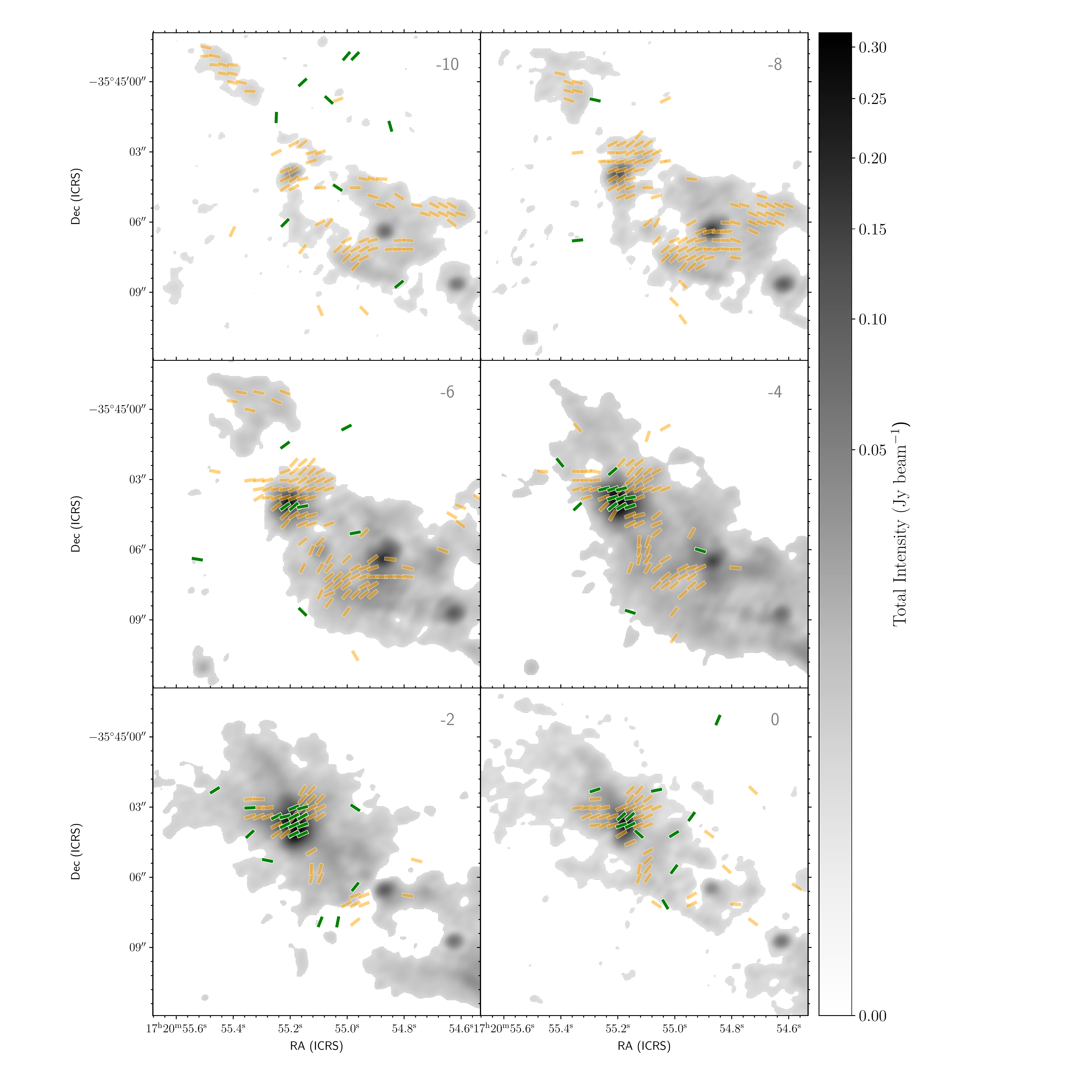}
\smallskip
\caption{Same as Figure \ref{fig:NGC6334INCS}, but here we show the C$^{33}$S$(J=5 \rightarrow 4)$ emission. The total intensity C$^{33}$S emission is shown as a gray-scale. The polarized emission from CS is shown in orange pseudo-vectors while the polarized emission from C$^{33}$S is shown in semi-transparent green pseudo-vectors. The significance of the polarization pseudo-vectors is 3$\sigma$ after debiasing where the average channel map noise level, before debiasing, is $\left< \sigma \right> = 770 \mu$Jybeam$^{-1}$.
\label{fig:NGC6334INC33S}
}
\bigskip
\end{figure*}

\subsection{The dispersion function analysis and the strength of the magnetic field}
\label{sse:disp}

\begin{figure*} 
\includegraphics[width=0.5\hsize]{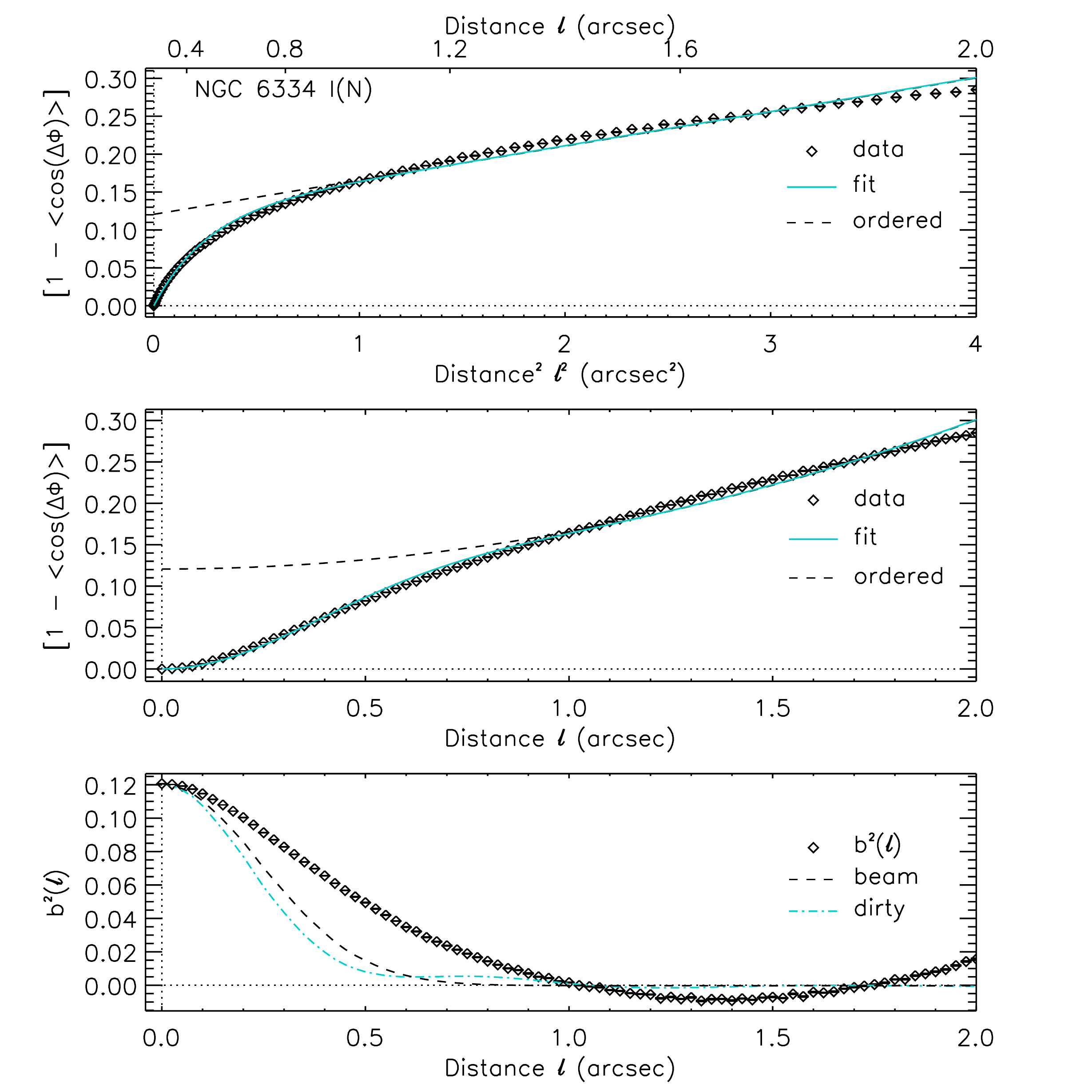}
\includegraphics[width=0.5\hsize]{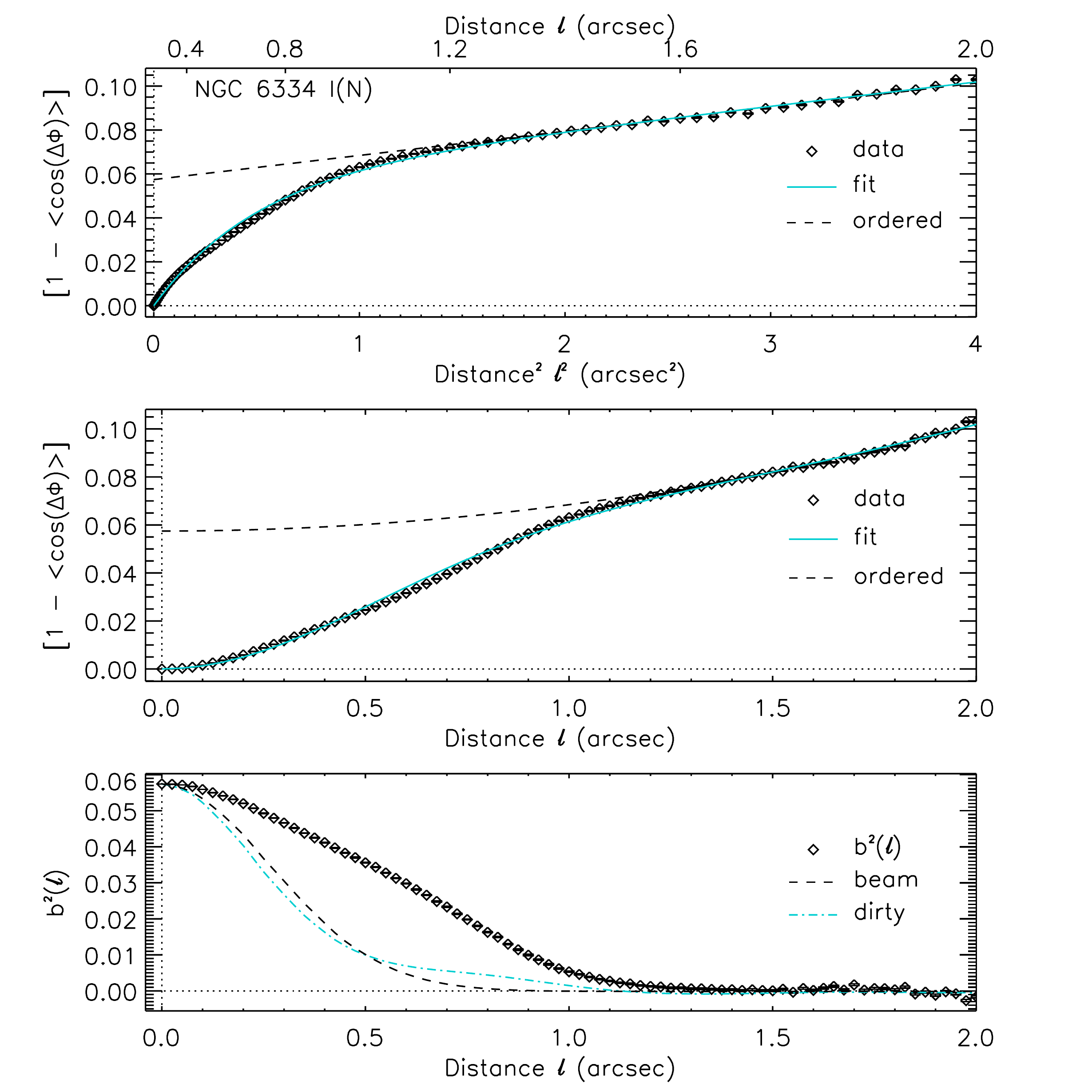}
\smallskip
\caption{Dispersion analysis of the ALMA NGC6334I(N) data. {\bf Left.} The dispersion function  calculated as $[1 - \left<\mathrm{cos( \Delta \phi ) } \right>]$ (in symbols) is plotted as a function of $l^{2}$ at the top. The ordered component is also shown using a broken curve. The least-squares fit of the Gaussian turbulence model is plotted in turquoise as a solid curve.  At the middle we also plot the dispersion function, but as a function of $l$. At the bottom, the signal-integrated turbulence autocorrelation function $b^{2}(l)$ (in symbols), along with the  autocorrelated Gaussian beam (segmented curve), and the ALMA dirty beam (solid turquoise curve) are plotted. From the fit to the data, we derive turbulence correlation length of $\delta = 0^{\prime \prime}.262 \pm 0^{\prime \prime}.008$, or $\sim$ 2 mpc, at the distance to NGC6334I(N), and $\left< B_{t}^{2} \right> / \left< B^{2} \right> = 0.29 \pm 0.01$  {\bf Right.} Same as the left panel, but using the integrated CS polarized emission data over the velocity interval -6 to -2 {\kms}. The analysis yields  
a turbulence correlation length of $\delta = 0^{\prime \prime}.42 \pm 0^{\prime \prime}.011$, or $\sim$ 2.6 mpc, and $\left< B_{t}^{2} \right> / \left< B^{2} \right> = 0.08 \pm 0.0013$.
\label{fig:disp}
}
\bigskip
\end{figure*}

The DCF method does not consider the effects of finite angular resolution or integration along the line of sight,
which at lower resolutions smooth out the polarized emission, reducing its dispersion, and therefore it might
 overestimates the field strength. Moreover, the effect of self-gravity in bending the field lines, which will affect the dispersion in the polarization position angle, is also not considered by DCF and its variants. Thus and because of the exquisite uv-coverage, resolution, and sensitivity of ALMA, local deviations in the position angle from the main field model, due to gravity, makes the applicability of the DCF method more challenging to the data.  
 Here, we discuss how the different DCF variants used in this work attempt to correct the polarization position angle dispersion.
 To account for poor resolution in polarization maps, \citet{Heitsch2001} used the geometric mean  between two modified DCF equations. The first modification attempts to address the small angle approximation by replacing the polarization angle dispersion by the dispersion of the tangent values of the position angle. The second modification attempts to deal with the case where the dispersion in the field lines is larger than the mean field. They do this by considering the 3-dimensional expansion of the field where all of the random components are assumed to be the same. This yields an equation also dependent on the dispersion of the tangent of the position angle values \citep[see equation 11 ][]{Heitsch2001}. Although consistent with their own simulations, their variant appears to underestimate the field strength when compared to other numerical results \citep{Falceta2008}. In our data, this method yielded an estimate which is a factor $\sim$ 10 smaller than the other estimates. It is likely that this is because local deviations from the ``hourglass'' main field morphology  produced larger values when using the tangent function, which  yields a larger dispersion and therefore a smaller estimate. In contrast, the modification proposed by \citet{Falceta2008} was implemented by assuming that  $\delta B/B$ is a global relation and thus they modified DCF by taking the tangent of the dispersion angle instead of $\delta \phi$ \citep[see equation 9 in ][]{Falceta2008}. This was proposed to address larger angle dispersions due to an increasing turbulent component in the field. However, this will rapidly decrease the  field strength as the tangent function quickly diverges for $\delta \phi > 60^{\circ}$. Recently, the DCF method was revisited and another correction was proposed which considers the compressible modes from small amplitude MHD waves (magnetosonic), instead of purely Alfven transverse waves \citep{Skalidis2021}. This method implements a substitution in the DCF equation in the form of $\delta \phi \rightarrow \sqrt{\delta \phi}$, which seem to improve the recovery of the field strength when applied to their numerical simulations.   Although it is not clear if the dominant mode perturbing the main field component is transverse (Alfvenic) or compressible (small amplitude magnetosonic waves), this variant will yield 
estimates which are smaller in value than the original DCF. The three previously described DCF variants used simulations to test the proposed modifications. All of the simulations assumed ideal MHD, which might not be representative of the physical conditions in NGC6334I(N). In fact, ideal MHD may produce artificially tangled magnetic field morphologies which will affect the polarization position angle dispersion, particularly in the line of sight. Furthermore and because of the low ionization rates in dense molecular clumps ($\sim 10^{-7}$), non-ideal MHD effects, such as ambipolar diffusion, are unavoidable \citep{Hennebelle2019}.
 
 The ADF method provides a way to quantify the turbulent component in the field and to better estimate the value of $\delta \phi$. This is done by fitting a structure function of the polarization position angle data, by incorporating  a turbulence model, and also by considering the effects of interferometric filtering
\citep[for a summary of the technique see Section 2 of][]{Houde2016}. This analysis allow us to associate $\delta \phi = \left[\left< B_{t}^{2} \right> / \left< B^{2} \right> \right]^{1/2}$, where the quantity on the right-hand side is the ratio of turbulent to total magnetic energy, one of the quantities derived from the dispersion-function analysis. We can then use $\delta \phi$ to estimate the field strength in the plane of the sky via the usual DCF technique, which takes the form of

\begin{equation}
    B_{\mathrm{pos}} \simeq \sqrt{4\pi\rho}\sigma(v) \left[ \frac{ \left< B_{t}^{2} \right> }{ \left< B^{2} \right> } \right]^{-1/2} \,\,,
\end{equation}

\noindent
where $\sigma(v)  = \Delta V / 2\sqrt{2\log{2}}$, $\Delta V$ is the FWHM line-width of the C$^{33}$S line, and $\rho$ is the volume density. 
By applying this analysis to the polarized dust emission data, considering the same region as before (see purple ellipse in Figure \ref{fig:NGC6334IN_POL}), we obtain a turbulence correlation length of $\delta = 0^{\prime \prime}.262 \pm 0^{\prime \prime}.008$, or $\sim$ 2 mpc, at the distance to NGC6334I(N), and $\left< B_{t}^{2} \right> / \left< B^{2} \right> = 0.23 \pm 0.01$, which we use to estimate a plane-of-sky magnetic field strength $B_{\mathrm{pos}}$ = 24 mG (see Figure \ref{fig:disp} and Table \ref{tab:B} for the results).
Furthermore, we apply the same analysis to the CS polarized emission data considering a wide velocity range, and obtain   a good fit between -6 to -2 {\kms}. In this interval, we find a turbulence correlation length of $\delta = 0^{\prime \prime}.42 \pm 0^{\prime \prime}.011$, or $\sim$ 2.6 mpc, and $\left< B_{t}^{2} \right> / \left< B^{2} \right> = 0.08 \pm 0.0013$, which yields a plane-of-sky magnetic field strength $B_{\mathrm{pos}}$ = 2.8 mG. Note, the analysis was applied to the integrated data between -6 to -2 {\kms}, but given the small differences in $\delta \phi$ within the interval, we used the same  $\left< B_{t}^{2} \right> / \left< B^{2} \right>$ value to estimate $B_{\mathrm{pos}}$ at each velocity channel (see Table \ref{tab:B}).
The differences in the field strength estimation is largely due to the differences in density used. That is, the dispersion analyses only contribute a factor of $\sim 1.7$ (lowering the dust estimate) while the densities increase the dust value by $\sim 14$.

Recent work by \cite{Liu2021} analyzed  ideal MHD simulations of proto-cluster formation at clump scales. They applied various statistical methods to synthetic magnetic field maps to study the applicability of the DCF method and the variants used here (including ADF). Because  the magnetic morphology in NGC6334I(N) has an ``hourglass'' shape, it is likely that we are in the strong-field regime. Thus,
\citeauthor{Liu2021} results suggests that the magnetic field strength estimates derived here are good to a factor of a few, again, subjected to the caveat of ideal MHD simulations.


We have five estimates for the field strength onto the plane of the sky. Each of them originate from modifications to the DCF method that try to 
address finite resolutions,  the polarization  angle dispersion value due to a random component of the field, and the effect of a different perturbation mode. Because none of these methods used here consider all of the  
relevant physics in this region, e.g. self-gravity,
the ``true'' value for the field strength remains unconstrained. We lack an actual measurement of the field strength such the one provided by the Zeeman effect.
Although still contested \citep[see ][]{Jiang_2020}, results from Zeeman measurements show that 
the field strength will grow with density as a power law, or $\sim n^{2/3}$ \citep{Crutcher2019}. Thus, the DCF and its variants still give us a first order statistical approximation to the ``true'' magnetic field strength onto the plane of the sky. We quantify a final estimate by taking the average of all five estimates obtaining $\left< \mathrm{B_{pos}} \right> = 16 $ mG
 as the average magnetic field onto the plane of the sky at densities of $n = 4.2 \times 10^{7}$ cm$^{-3}$, and $\left< \mathrm{B_{pos}} \right> = 2$  mG at densities of $n = 2.0 \times 10^{5}$ cm$^{-3}$ when considering the polarized CS emission.

\subsection{Comparison with other HMSFR}

By  comparing our results from NGC6334I(N) to other HMSFR,  we  seek to discover if  there is a pattern in  the  physical  conditions  of   regions where the magnetic field has a  clear and distinctive  shape,  such  as an  ``hourglass''   morphology. 
Examples of similar magnetic field morphologies to  NGC6334I(N) are  cores like G240 where the ``hourglass''  magnetic  field,  appears  as a ``textbook''  case  for magnetic controlled star formation  with  a bipolar  outflow closely  aligned  to  both the rotation and  magnetic field axes  \citep{Qiu2014}.  Note, the G240 mass, 95 {\Msun}, is substantially larger than the  $\sim$ 26 {\Msun} of 
the combined 1b and 1c core masses in NGC6334I(N) as derived from our data inside the purple oval region shown in Figure \ref{fig:NGC6334IN_POL}. 
Another  example is  the massive core G31.41, with a mass comparable to  NGC6334I(N) \citep[about  26 {\Msun} from ][]{Beltran2019},  has also been shown to exhibit an ``hourglass''   magnetic  field morphology  from core \citep{Girart2009} to envelope scales \citep{Beltran2019}. However,  the alignment between the outflow,  rotation, and magnetic axes is less clear here when compared to G240. The length-scales where the ``hourglass'' shape is traced in these two  sources are   similar  to what we see in NGC6334I(N).
For instance, in G31.41 the field morphology is seen preserved through a scale range that matches the lower end in the NGC6334I(N) scales. The data obtained from G240 traces  the field at the core scales also  where the ``hourglass'' is seen in NGC6334I(N).
However, not all HMSFRs show ``hourglass''  magnetic field morphologies.
For instance, in the W43-Main molecular complex, the  W43-MM1 \citep{Cortes2016,Arce2020} and W43-MM2 clumps \citep{Cortes2019} exhibit magnetic field morphologies that are primarily radial over their most massive cores, which is expected when gravity dominates the dynamics. 
  W43-Main harbors some of the most massive 
protostars currently known \citep[$\gtrsim 100$ {\Msun},][]{Cortes2016,Motte2018a,Cortes2019}, whereas the mass of the central cores in NGC6334I(N) are only about  $\sim 26$ {\Msun} in total when considering our data.  However and because of   the angular scales sampled by our ALMA data, we might be missing flux  which  might make NGC6334I(N) appear less massive than other  regions. Nonetheless, this difference in core mass is not significant when  comparing W43-Main with G240 where the core masses are comparable, but the field shapes are completely different. 
A totally  different magnetic field  morphology is  seen  in IRAS 180089-1732 where the field was found to have an spiral morphology  \citep{Sanhueza2021}.
Previous mapping of this  source at clump scales appears to show  the same field pattern \citep{Beuther2010} as seen by  ALMA at envelope scales. In this case, the total core mass is estimated to be 75 {\Msun} from the ALMA data,  which is also comparable to G240  and  in the  lower range  from the W43-Main  estimates.
Thus,  it is also uncertain whether the core mass is a decisive  factor to explain the differences in the field shape seen across these HMSFRs.

High mass star forming  cores are usually  surrounded  by  {\HII} regions which  provide significant  radiative feedback.
It is possible that  radiation   pressure  coming from {\HII} regions may compress the  field in conjunction with the effects of gravity, which  the field may resist if strong enough \citep[e.g. see ][ for an example in the Carina nebula]{Li2006,Shariff2019}.
For  instance, W43-Main is part of a giant molecular complex which has at its center a large {\HII} region powered by a
number of O7 Wolf-Rayet stars, which appear to be not only ionizing the boundaries of W43-Main but also compressing the gas \citep{Blum1999,Motte2003}, while NGC6334 contains a group of smaller {\HII} regions  known as the ``Cat's Paw''
which seem distributed along the filament \citep{Russeil2016}. This  also seems to be the  case  for G31.41, which is surrounded by  both compact and extended {\HII} regions 
(J. M. Girart private communication). In contrast, for G240  and IRAS 180089-1732, the situation seems unclear as the cores appear  to be more isolated than   NGC6334I(N), G31.41, and W43-Main. Although we note these differences,
in this simple analysis we are certainly ignoring a number of other factors such as chemical diversity, stage of evolution, possible initial conditions, among many others. 
Thus, acquiring sufficient statistical cases is paramount to increase our understanding  about how stars form in high mass star forming  regions   and what is the role of the magnetic field. As part of this MagMaR project, we have acquired a comprehensive sample that is sufficiently large to allow us to begin addressing these questions in future work.

\section{SUMMARY AND CONCLUSIONS}\label{se:conc}

We present ALMA observations of polarized dust, CS$(J=5 \rightarrow 4)$, and C$^{33}$S$(J=5 \rightarrow 4)$ emission towards NGC6334I(N). From these data we find:

\begin{itemize}
    \item The magnetic field derived from the ALMA polarized dust emission data shows a clear ``hourglass'' morphology over the 1b, 1c, and possibly 1a cores. This shape is preserved from clump to envelope scales when considering both the SMA and JCMT data. 
    \item We obtained polarized emission  from CS and C$^{33}$S $J=5\rightarrow4$ transition. We modelled the total intensity for both the CS and C$^{33}$S lines using the MADCUBA software; we calculate optical depths of 32 and 0.1 for each line, respectively.
    \item The polarized emission from CS nicely traces the same magnetic field ``hourglass'' morphology  inferred from the polarized dust emission within the -12 to 10 ${\kms}$ velocity range. We estimated  a number density  of 2 $\times 10^{5}$ cm$^{-3}$ as traced by the CS emission; 2 order of magnitude less than the $4.2\times10^{7}$  cm$^{-3}$ derived from dust emission. This allow us to obtain a tomographic view of the field in this region from a single dataset.
    \item We also report polarized emission from C$^{33}$S. While there are fewer independent detections of polarization, the polarized emission appears consistent with the CS results.
    \item We estimate the magnetic field strength onto the plane of the sky from both the dust and the CS data by using multiple methods. We obtain an average field strength estimate of $\left< \mathrm{B}_{\mathrm{pos}} \right> = 16$ mG from the dust and $\left< \mathrm{B}_{\mathrm{pos}} \right> \sim 2$ mG from the CS emission, when considering the -6 to -2 {\kms} velocity range.
\end{itemize}

\newpage
\appendix

\section{Stokes $V$ velocity channel maps}
\label{apx1}

\begin{figure*} 
\includegraphics[width=1.0\hsize]{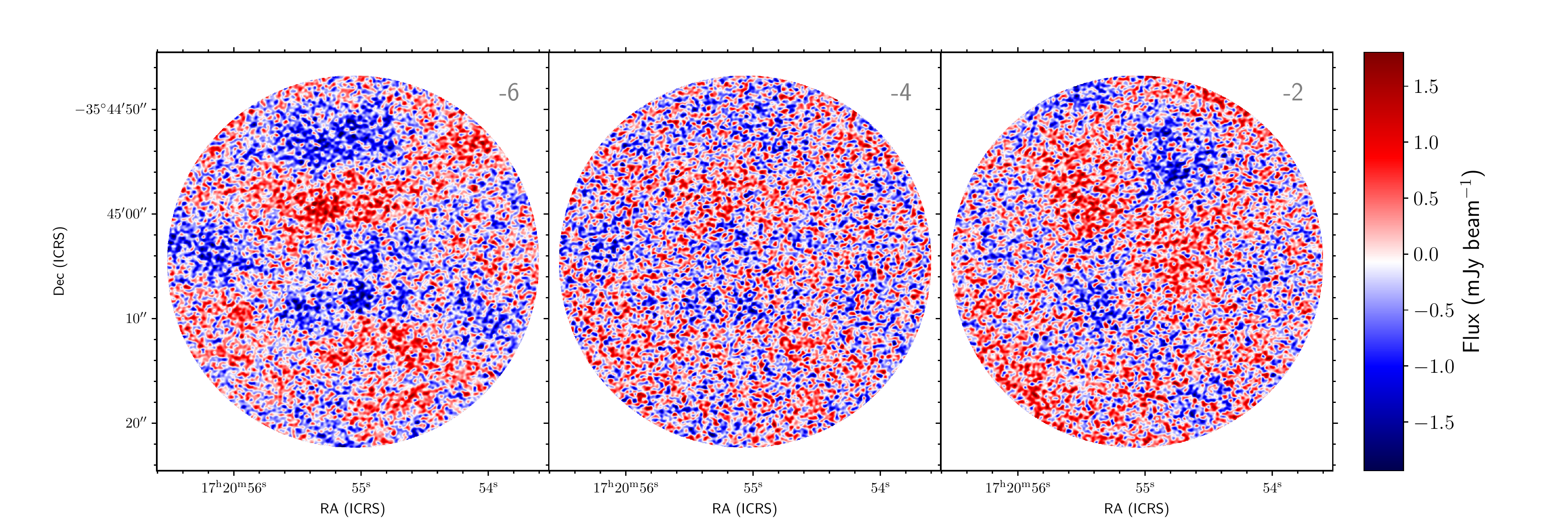}
\smallskip
\caption{The velocity  channel maps for the flat, not primary beam corrected, Stokes $V$ are shown here from -6 to -2 {\kms} using a divergent color scheme to indicate both negative and positive values. The velocity of  each channel is indicated in the top right corner at each map.
\label{fig:StokesV}
}
\bigskip
\end{figure*}

\begin{figure*} 
\includegraphics[width=1.0\hsize]{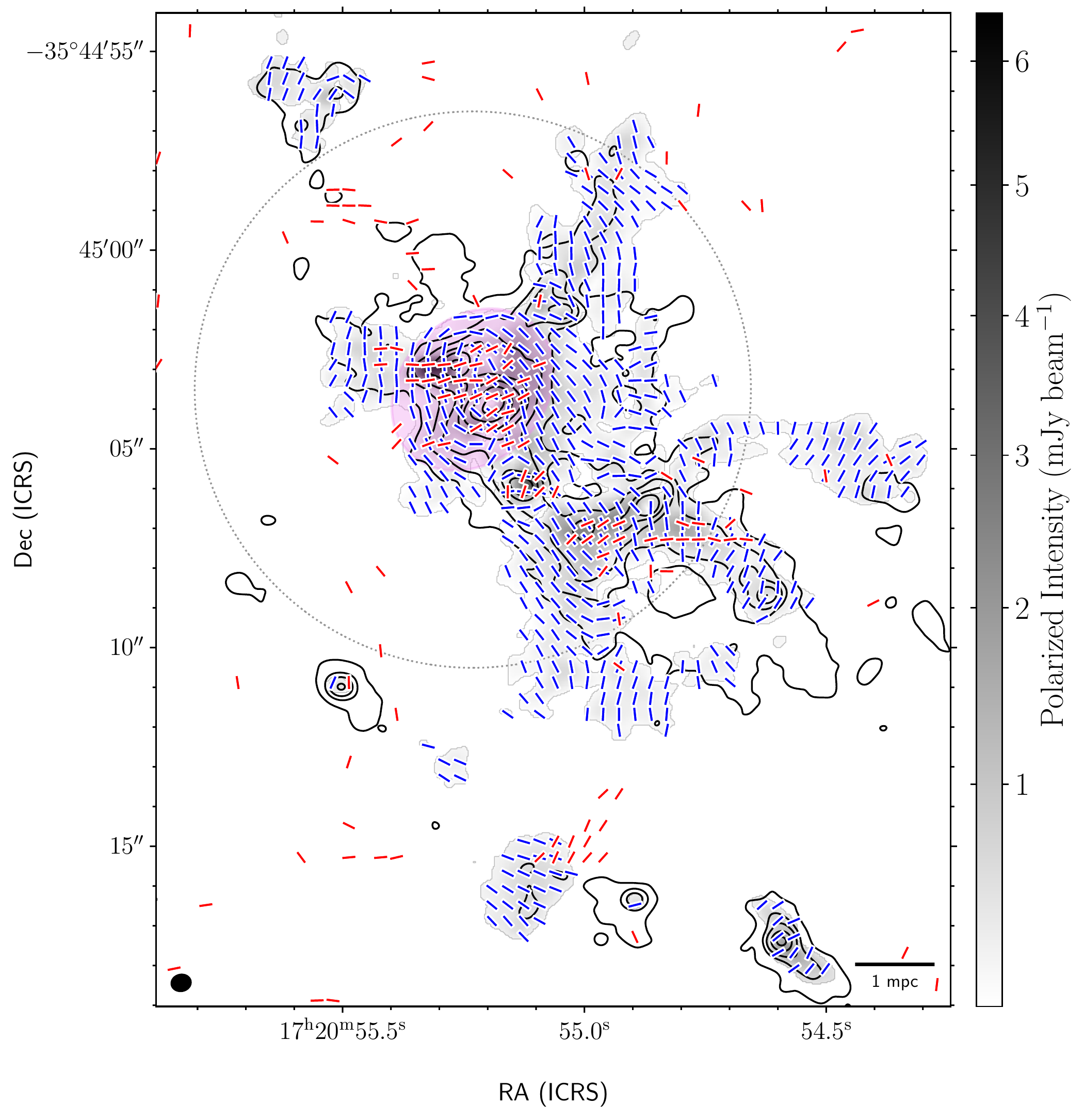}
\smallskip
\caption{Here we show un-debiased polarization maps from both dust (blue) and CS (red) emission. The Figure follows the same layout at Figure \ref{fig:NGC6334IN_POL}. The CS polarization maps was produced by imaging a continuum substracted, u-v averaged, measurement sets. The first session was uv-averaged between 244.919 to 244.968 GHz while the second session was uv-averaged between 244.935 to 244.984 GHz. 
\label{fig:mom0}
}
\end{figure*}

Figure \ref{fig:StokesV} shows the Stokes $V$ velocity channel maps for the range between -6 to -2 {\kms} which we used to estimate the magnetic field strength onto the plane of the sky.  The maps statistics are given in Table \ref{tab:stokesStats}. Although there seem to be some structure  in the maps, this structure is not consistent with the compact linear polarization that we see from CS and what we would expect if ARS were happening in NGC6334I(N). This structure is likely off-axis instrumental leakage which is under the current ALMA accuracy \citep{Hull2020b,Cortes2021b}.
Additionally, we show the integrated CS emission map superposed to the polarized dust emission; note, not the inferred magnetic field onto the plane of the sky (see Figure \ref{fig:mom0}). 
We are showing the unbiased data to show the full extent of the data. 
The integrated CS emission map was averaged in uv-space over the range of the line and both polarization pattern appear to be orthogonal to each other as expected based on the argumentation exposed in this work (see section \ref{se:discussion}).

\section{The normalization problem}
\label{apx2}

It was recently discovered that strong molecular
line emission can be detected in the auto-correlations of ALMA data, thereby biasing the
spectrally-resolved normalization, and sometimes the nominally
off-source Tsys calibration measurements detect unrelated astronomical
line emission.
This problem will affect the calibrated fluxes obtained from molecular line emission biasing the determination of any astrophysical quantity derived from such fluxes. A re-normalization strategy was put in place by ALMA
which corrects for this effect including the polarization data reported here \citep{Moellenbrock2021}.
We  detect the CS$(J=5\rightarrow4)$ line only in the auto-correlations and not in the T$_{sys}$ spectra. By applying the aforementioned ALMA correction, we estimate that the error introduced in $\sigma_{\phi} < 0.5^{\circ}$ for the polarization position
angle and $\sigma_{Pfrac} = 0.01 \%$, which are essentially negligible for the scope of this work. Although we find small errors in these quantities, we do find noticeable differences in the Stokes spectra. The difference is because the polarization position angle and the fractional polarization are derived from ratios of the Stokes parameter which cancel out the normalization effect. Therefore, data that requires modeling of the Stokes parameters (e.g. Zeeman measurements), should carefully apply this correction and inspect the spectra appropriately.

\bigskip

\begin{deluxetable*}{c c c c c c c c c c c c c}        
\tablecolumns{13}
\tablewidth{0pt}
\tablecaption{Magnetic Field Strength Estimations\label{tab:B}}
\tablehead{
    \colhead{Tracer} &
    \colhead{Velocity} &
    \colhead{N\tablenotemark{\scriptsize a}}&
    \colhead{$n$} &
    \colhead{$\left< \phi \right>$\tablenotemark{\scriptsize b}} &
    \colhead{$\delta \phi$\tablenotemark{\scriptsize b}} &
    \colhead{$\Delta$ V} &
    \colhead{B$_{1}$\tablenotemark{\scriptsize c}} &
    \colhead{B$_{2}$\tablenotemark{\scriptsize d}} &
    \colhead{B$_{3}$\tablenotemark{\scriptsize e}} &
    \colhead{B$_{4}$\tablenotemark{\scriptsize f}} &
    \colhead{B$_{5}$\tablenotemark{\scriptsize g}} &
    \colhead{${\left< B_{t} \right>}^{2}/{\left< B \right>}^{2}$} \\
    \colhead{ } &
    \colhead{[{\kms}]} & 
    \colhead{[$10^{24}$ cm$^{-2}$]} &
    \colhead{[$10^{5}$ cm$^{-3}$]} &
    \colhead{[$^{\circ}$]} &
    \colhead{[$^{\circ}$]} &
    \colhead{[{\kms}]} &
    \colhead{[mG]} &
    \colhead{[mG]} &
    \colhead{[mG]} &
    \colhead{[mG]} &
    \colhead{[mG]} & 
    \colhead{} 
}
\startdata
Dust & - & 2.0 & 423.9 & 29.3 & 29.4 & 5.3 & 21.9 & 19.9 & 1.4 & 23.6 & 11.1 & 0.23 $\pm$ 0.01\\
CS & -6.0 & 2.0 & 2.0 & 179.3 & 6.7 & 5.3 & 3.24 & 3.18 & 0.060 & 2.75 & 1.58 & 0.08 $\pm$ 0.001\\
CS & -4.0 & 2.0 & 2.0 & 179.2 & 6.5 & 5.3 & 2.49 & 2.40 & 0.116 & 2.75 & 1.39 & 0.08 $\pm$ 0.001\\
CS & -2.0 & 2.0 & 2.0 & 179.3 & 6.2 & 5.3 & 2.39 & 2.31 & 0.165 & 2.75 & 1.36 & 0.08 $\pm$ 0.001\\
\enddata
\tablenotetext{a}{The column density corresponds to the region
used to extract $\delta \phi$.}
\tablenotetext{b}{$\left< \phi \right>$ is the average polarization position angle (PA) and $\delta \phi$ is the PA dispersion
(calculated using circular statistics).}
\tablenotetext{c}{Estimations of the magnetic field, in the plane of the sky, done with the original DCF method.}
\tablenotetext{d}{Estimations of the magnetic field in the plane of the sky, done using the corrections implemented by Equation 9 in \citet{Falceta2008}.}
\tablenotetext{e}{Estimations of the magnetic field in the plane of the sky, done using the corrections implemented by Equation 12 in \citet{Heitsch2001}.}
\tablenotetext{f}{Magnetic field strength estimated by using the ratio of turbulent to total magnetic energy.}
\tablenotetext{g}{Magnetic field strength estimated by using the DCF modification proposed by \citet{Skalidis2021}.}
\end{deluxetable*}

\textit{Facilities:} ALMA.

\textit{Software:} APLpy, an open-source plotting package for Python hosted at \url{http://aplpy.github.com} \citep{Robitaille2012}.  CASA \citep{McMullin2007}.  Astropy \citep{Astropy2018}. MADCUBA \citep{Martin2019}.

\acknowledgements
P.C.C. would like to thank  A. Sanchez-Monge and  M. Sadaghiani for providing machine readable tables from their data.
P.C.C acknowledges publication support from ALMA and NRAO.
P.S. was partially supported by a Grant-in-Aid for Scientific Research (KAKENHI Number 18H01259) of the Japan Society for the Promotion of Science (JSPS).
M.H.'s research is funded through the Natural Sciences and Engineering Research Council of Canada Discovery Grant RGPIN-2016-04460.
C.L.H.H. acknowledges the support of the NAOJ Fellowship and JSPS KAKENHI grant 20K14527.
C.L.H.H. and J.M.G. acknowledge the support of JSPS KAKENHI grant 18K13586.
J.M.G. acknowledges the support of the Spanish grant AYA2017-84390-C2-R (AEI/FEDER, UE).
L.A.Z acknowledges financial support from CONACyT-280775 and UNAM-PAPIITIN110618 grants, Mexico.
This paper makes use of the following ALMA data: 2018.1.00105.S.
ALMA is a partnership of ESO (representing its member states), NSF (USA) and NINS (Japan), together with NRC (Canada), MOST and ASIAA (Taiwan), and KASI (Republic of Korea), in cooperation with the Republic of Chile. The Joint ALMA Observatory is operated by ESO, AUI/NRAO and NAOJ.
The National Radio Astronomy Observatory is a facility of the National Science Foundation operated under cooperative agreement by Associated Universities, Inc.


\bibliography{biblio}{}
\bibliographystyle{aasjournal}

\end{document}